\documentclass[conference]{IEEEtran}

\usepackage{color}
\usepackage{comment}
\usepackage{dsfont}
\usepackage{cite}
\usepackage{amsmath,amssymb,amsfonts}
\usepackage{algorithmic}
\usepackage{graphicx}
\usepackage{textcomp}
\usepackage{cite}
\usepackage{amsmath,amssymb,amsfonts}
\usepackage{algorithmic}
\usepackage{graphicx}
\usepackage{epstopdf}
\usepackage{subfigure}
 \graphicspath{{\figures}}
\usepackage{textcomp}
\usepackage{xcolor}
\usepackage{xpatch}
\makeatletter
\def\changeBibColor#1{
\in@{#1}{flightenergy,channel,UAV3,UAV2,Mappo,qtran,qmix,sinr5,sinr4,sinr3,sinr1,reflect3,twostep2,twostep,reflect4,reflect2,reflect1,reflect,via,aoa2,rssi,TWCaccuracy,addadd1,addadd2,addadd3,addadd4,addadd5,addadd6}
\ifin@\color{black}\else\normalcolor\fi
}
\xpatchcmd\@bibitem
{\item}
{\changeBibColor{#1}\item}
{}{\fail}
\xpatchcmd\@bibitem
{\item}
{\changeBibColor{#1}\item}
{}{\fail}
\makeatother

\usepackage{verbatim}
\usepackage{makecell}
\usepackage{booktabs}
\usepackage{CJK}
\usepackage{multirow}

\usepackage{algorithm}
\usepackage{algorithmic}

\makeatletter
\newcommand{\removelatexerror}{\let\@latex@error\@gobble}
\makeatother
\IEEEoverridecommandlockouts 
\begin{document}
\begin{CJK}{UTF8}{gbsn}
\title{
Collaborative Reinforcement Learning Based Unmanned Aerial Vehicle (UAV) Trajectory Design for 3D UAV Tracking}
\author{
Yujiao Zhu, \emph{Student Member, IEEE}, Mingzhe Chen, \emph{Member, IEEE},\\ Sihua Wang, \emph{Student Member, IEEE}, Ye Hu, \emph{Member, IEEE},\\ Yuchen Liu, \emph{Member, IEEE}, and Changchuan Yin, \emph{Senior Member, IEEE}\\
\thanks{Y. Zhu, S. Wang, and C. Yin are with the Beijing Laboratory of Advanced Information Network, Beijing University of Posts and Telecommunications, Beijing, 100876, China (E-mail: yjzhu@bupt.edu.cn; sihuawang@bupt.edu.cn; ccyin@bupt.edu.cn).

M. Chen is with the Department of Electrical and Computer Engineering and Institute for Data Science and Computing,
University of Miami, Coral Gables, FL, 33146, USA (Email: mingzhe.chen@miami.edu).

Y. Hu is with the Department of Industrial and System Engineering,
University of Miami, Coral Gables, FL, 33146, USA (Email: yehu@miami.edu).

Y. Liu is with the Department of Computer Science, North Carolina State University, Raleigh, NC, 27695, USA (Email: yuchen.liu@ncsu.edu).

A preliminary version of this work \cite{conference} is accepted in the Proceedings of the 2023 IEEE International Global Communications Conference (GLOBECOM).
}
}
\maketitle
\vspace{-4em}
\begin{abstract}
In this paper, the problem of using one active unmanned aerial vehicle (UAV) and four passive UAVs to localize a 3D target UAV in real time is investigated. In the considered model, 
each passive UAV receives reflection signals from the target UAV, which are initially transmitted by the active UAV. The received reflection signals allow each passive UAV to estimate the signal transmission distance which will be transmitted to a base station (BS) for the estimation of the position of the target UAV.
Due to the movement of the target UAV, each active/passive UAV must optimize its trajectory to continuously localize the target UAV. 
Meanwhile, since the accuracy of the distance estimation depends on the signal-to-noise ratio of the transmission signals, the active UAV must optimize its transmit power.
This problem is formulated as an optimization problem whose goal is to jointly optimize the transmit power of the active UAV and trajectories of both active and passive UAVs so as to maximize the target UAV positioning accuracy. To solve this problem, a Z function decomposition based reinforcement learning (ZD-RL) method is proposed. 
Compared to value function decomposition based RL (VD-RL), the proposed method can find the probability distribution of the sum of future rewards to accurately estimate the expected value of the sum of future rewards thus finding better transmit power of the active UAV and trajectories for both active and passive UAVs and improving target UAV positioning accuracy. Simulation results show that the proposed ZD-RL method can reduce the positioning errors by up to 39.4\% and 64.6\%, compared to VD-RL and independent deep RL methods, respectively.

\end{abstract}


\section{Introduction}
Unmanned aerial vehicle (UAV) localization has gained significant attention from academic and commercial fields since it supports a wide range of applications in military, assistance and industrial scenarios~\cite{UAVattack,tut,ZHY,addadd1}.
For example, when UAVs perform attack missions in the military field, it is necessary to locate and track unauthorized UAVs in real time \cite{QQW,addadd2}.
However, achieving accurate UAV positioning faces several challenges. 
First, UAVs are moving at a high speed, and thus estimating the real-time positions of UAVs is challenging.
Second, since the coordinates of UAVs are three-dimensional (3D), estimating 3D coordinates of UAVs requires more sensors (at least four sensors) and complex positioning algorithms.
Third, dynamic wireless environments such as electromagnetic interference, transmit power allocation, and available communication resources will affect the transmission of pilot signals used for UAV localization thus affecting UAV localization accuracy~\cite{ZHY2,newadd,addadd6}. 

\subsection{Related Works}
 Recently, several existing works such as \cite{single1,RGBD,Ismail2,3,5,aoa2,rssi} have focused on UAV localization. The authors in \cite{single1} and \cite{RGBD} considered the use of a single camera sensor 
 to track movement of UAVs. However, the positioning algorithms used in \cite{single1} and \cite{RGBD} must be implemented based on unique hardware and high computational resource. 
 \textcolor{black}{The authors in \cite{Ismail2,3,5,aoa2,rssi} used radio-frequency (RF) signals to estimate the positions of UAVs. 
 In particular, in \cite{Ismail2,3}, the authors obtained the arrival time of transmitted signals from several sensors and determined the 3D positions of UAVs. The authors in \cite{5} jointly used the arrival angle and departure angle of transmitted signals to estimate the positions of UAVs thus reducing the number of sensors used for UAV localization. The authors in \cite{aoa2} studied the UAV trajectory optimization problem and estimate the position of the UAV based on angle information of arrival signals. The authors in \cite{rssi} used the received signals strength to measure distance information and analyzed the impact of different distance measurement errors on UAV localization performance.
However, the authors in \cite{single1,RGBD,Ismail2,3,5,aoa2,rssi} did not consider how the positions of sensors affect the UAV localization accuracy and they also did not consider the optimization of the deployment of sensors. In fact, the positions of sensors will significantly affect the UAV positioning accuracy \cite{deployment}.  Meanwhile, most of these works \cite{single1,RGBD,Ismail2,3,5,aoa2,rssi} assumed that the values of signal-to-noise ratio (SNR) of transmitted signals are constant, which is impractical in actual wireless networks. 
In addition, most of these works \cite{single1,RGBD,Ismail2,3,5,aoa2,rssi} assumed that a central controller knows the positions of all sensors and channel state information (CSI) in advance such that the central controller will directly use this information for UAV positioning. Therefore, these works \cite{single1,RGBD,Ismail2,3,5,aoa2,rssi} cannot be used for scenarios where the central controller cannot obtain the positions of sensors or CSI. }

Recently, a number of existing works \cite{UAVPositioning,Robust,Autonomous,sim-to-real,POMDP} have studied the use of reinforcement learning (RL) \cite{RVA} for UAV localization in the networks where the central controller cannot obtain all the information needed for UAV localization.
In particular, the authors in \cite{UAVPositioning} selected different ground sensors to optimize the UAV localization performance using a double deep Q-network based RL method. The authors in \cite{Robust} developed a domain randomization based RL algorithm and estimated the real-time position of a UAV using a monocular camera while considering environmental impacts such as wind gusts.
The authors in \cite{Autonomous} used time difference of signal arrival information measured by ground sensors to estimate 3D coordinates of UAVs and applied deep deterministic policy gradient (DDPG) and soft actor-critic methods to optimize Taylor series linearized localization approach. 
The authors in \cite{sim-to-real} analyzed the effects of measurement uncertainty on the performance of UAV localization based on a proximal policy optimization (PPO) algorithm in an environment with dynamic noise.
In \cite{POMDP}, the authors mapped UAVs' initial sensory measurements into control signals for localization and navigation by an actor-critic based deep reinforcement learning (DRL) algorithm.
However, the central controller in these works \cite{UAVPositioning,Robust,Autonomous,sim-to-real,POMDP} must collect sensing data from all sensors to determine the UAV movement, which will increase the communication overhead and the time used for UAV localization. Meanwhile, most of these works \cite{Robust,Autonomous,sim-to-real,POMDP} considered the use of statically installed sensors for UAV localization, which may not be used for localizing a UAV with a high movement speed.  
 \subsection{Contributions}
The main contribution of this work is to design a novel framework that can real-time monitor the position of a target UAV by controlled UAVs including four passive UAVs and one active UAV. The main contributions include:
\begin{itemize}
     \item[$\bullet$] We propose a UAV-based localization system to estimate the positions of the target UAV in which the active UAV transmits signals to the target UAV, while four passive UAVs collect the arrival time of signals transmitted from the active UAV to the target UAV, and then from the target UAV to passive UAVs. Next, each passive UAV estimates the distance from the active UAV to the target UAV, and then to the passive UAV. Such distance information is transmitted to the BS, which calculates the position of the target UAV. 
    \item[$\bullet$] In the considered UAV localization system, since the target UAV will change its position according to its performed task, each controlled UAV must optimize its trajectory to accurately localize the target UAV. Meanwhile, the accuracy of the distance information estimated by passive UAVs depends on the SNR of the signals transmitted from the active UAV and hence the active UAV must optimize its transmit power according to the movements of the target UAV and passive UAVs. This problem is formulated as an optimization problem that aims to maximize the localization accuracy of the target UAV via optimizing the transmit power of the active UAV and the trajectories of the active and passive UAVs. 
    \item[$\bullet$] To solve this problem, we propose a Z function decomposition based reinforcement learning (ZD-RL) method that enables each controlled UAV to determine its trajectory and the active UAV to determine its transmit power via its individual observation. Compared to value function decomposition methods \cite{VDN}, the Z function decomposition can find the probability distribution of the sum of future rewards such that each controlled UAV can accurately estimate the expected value of the sum of future rewards to update the parameters of its deep neural networks (DNNs). Hence, the proposed ZD-RL method can improve the efficiency and stability of optimizing the transmit power of the active UAV and the trajectories of controlled UAVs to minimize the positioning error of the target UAV. 
    \item[$\bullet$] To further minimize the positioning error of the target UAV, we analyze how the positions of the controlled UAVs affect the positioning error of the target UAV. Our analytical results show that the minimum positioning error of the target UAV can be achieved when the distance between each controlled UAV and the target UAV is minimized. 
\end{itemize}

Simulation results show that the proposed ZD-RL method can achieve up to 39.4\% and 64.6\% reduction in the positioning error of the positions of the target UAV compared to traditional value function decomposition based RL (VD-RL) and independent DRL methods, respectively. \emph{To the best of our knowledge, this is the first work that presents a UAV localization framework that 
utilizes one active UAV and four passive UAVs for 3D UAV positioning.}

The rest of this paper is organized as follows. The system model and problem formulation are described in Section \uppercase\expandafter{\romannumeral2}. The Z function decomposition based power allocation and trajectory design method is discussed in Section \uppercase\expandafter{\romannumeral3}. 
The optimal deployment of controlled UAVs for target UAV localization are analyzed in Section \uppercase\expandafter{\romannumeral4}.
In Section \uppercase\expandafter{\romannumeral5}, numerical simulation results are presented and analyzed. Finally, conclusions are drawn in Section \uppercase\expandafter{\romannumeral6}.
\begin{figure}[t!]
 \setlength\abovecaptionskip{0.cm}
\setlength\belowcaptionskip{-0.25cm}
\centering 
\includegraphics[width=8cm]{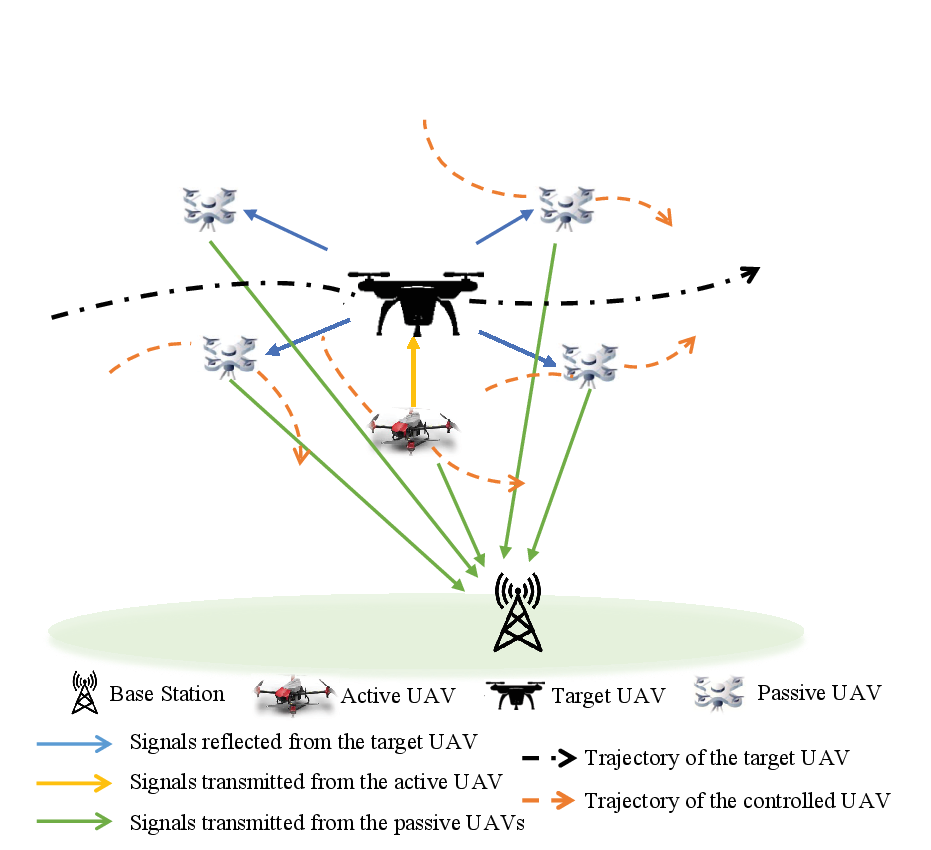}
\caption{Illustration of the considered UAV localization network.}
\label{scenario} 
\vspace{-0.6cm}
\end{figure}
\section{System Model and Problem Formulation}

\begin{figure}[t]
 \setlength\abovecaptionskip{0.cm}
\setlength\belowcaptionskip{-0.3cm}
\centering
\includegraphics[width=6.5cm]{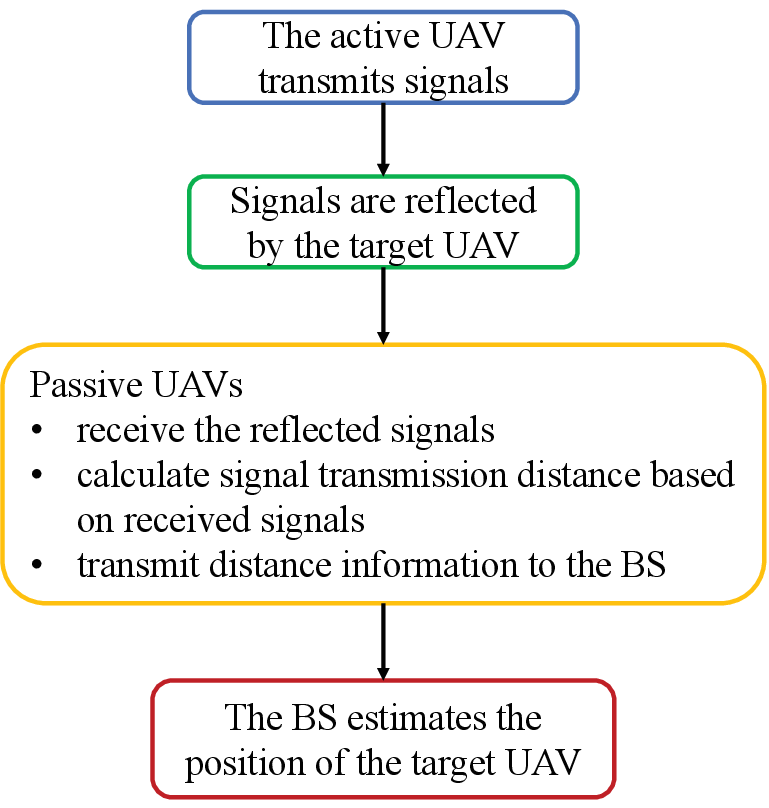}
\caption{\color{black}The flow chart of the considered UAV positioning process.}
\label{estimation}
\vspace{-0.6cm}
\end{figure}

Consider a UAV-assisted positioning network in which a ground BS and a set $\mathcal{M}$ of five controlled UAVs jointly monitor the position of the target UAV in real time, as shown in Fig. \ref{scenario}. 
The controlled UAVs consist of an active UAV and four passive UAVs\textcolor{black}{\footnote{\textcolor{black}{Since we use the traditional time difference of arrival (TDOA) method to calculate the three-dimensional (3D) coordinate of the target UAV \cite{TSML}, four passive UAVs are required to estimate the four signal transmission distances and calculate the 3D position of the target UAV.}}}.  \textcolor{black}{Here, the target UAV cannot directly transmit its position to the BS since the target UAV may not know its current position, or the target UAV may be an adversarial UAV and it will not share its position to the BS and passive UAVs.}
In our model, the active UAV first transmits signals to the target UAV which will reflect the signals to passive UAVs.  Then, passive UAVs estimate the signal transmission distance from the active UAV to the target UAV, and then to passive UAVs. The estimated signal transmission distance will be transmitted to the BS to calculate the position of the target UAV.
We assume that the real-time 3D coordinates of the controlled UAVs are known to the BS. \textcolor{black}{The flow chart of estimating the target UAV's position is shown in Fig. \ref{estimation}.}  Next, we first introduce the movement model of the active and passive UAVs. Then, the transmission links among the active UAV, target UAV, passive UAVs, and the BS are introduced. Finally, the positioning model and the optimization problem is formulated.



Let $\boldsymbol{u}_{m,t}=\left[x_{m,t},y_{m,t},z_{m,t}\right]^T$ be the 3D coordinate of UAV $m$ at time slot $t$. 
Hereinafter, we use a sequence number $0$ to represent the active UAV and a sequence number from $1$ to $4$ to represent a passive UAV. For example, $\boldsymbol{u}_{0,t}$ represents the coordinate of the active UAV and $\boldsymbol{u}_{m,t}$ with $1\leqslant m\leqslant 4$ is the coordinate of a passive UAV. Then, the coordinate of UAV $m$ is
\begin{equation}\label{passive motion}
\boldsymbol{u}_{m,t+1}\left(\phi_{m,t},\varphi_{m,t}\right)=\boldsymbol{u}_{m,t}+v_{m,t}\Delta_t\begin{bmatrix} \cos\varphi_{m,t}\cos\phi_{m,t}\\\sin\varphi_{m,t}\cos\phi_{m,t}\\\sin\phi_{m,t}
  \end{bmatrix},
\end{equation}
where $\varphi_{m,t}$ is the yaw angle, $\phi_{m,t}$ is the pitch angle, $v_{m,t}$ is the flight speed, and $\Delta_t$ is the time duration of a time slot.

\begin{table}[t!]
\caption{\label{parameters}List of Notations}  
\centering 
\setlength{\abovecaptionskip}{0.cm}
\setlength{\tabcolsep}{0.8mm}{
\begin{tabular}{|c|c|}  
\hline  
$\mathbf{Notation}$ & $\mathbf{Description}$ \\
\hline  
$M$ & Number of controlled UAVs \\
\hline  
$\boldsymbol{u}_{m,t}$ & Position of controlled UAV $m$ \\ 
\hline 
$v_{m,t}$ & Flight speed of controlled UAV $m$ \\
\hline  
$\Delta_t$ & Time duration of a time slot \\ 
\hline 
$\varphi_{m,t}$ & Yaw angle of controlled UAV $m$ \\
\hline  
$\phi_{m,t}$ & Pitch angle of controlled UAV $m$ \\ 
\hline  
$\tau_{m,t}$ & Transmit time of signals \\
\hline  
$c$ & Speed of light  \\ 
\hline  
$\boldsymbol{s}_t$ & Position of the target UAV \\
\hline  
$d_{m,t}$ & Distance from the target UAV to controlled UAV $m$ \\ 
\hline  
$p_{m,t}$ & Transmit power of controlled UAV $m$ \\
\hline  
$\omega_{m,t}$ & Random Gaussian noise \\
\hline  
$a_{t}$ & Transmitting signal \\
\hline  
$y_{m,t}$ & Received signals at passive UAV $m$ \\ 
\hline  
$x_{m,t}$ & Scattering coefficient of the target UAV \\
\hline  
$h_{m,t}$ & Path loss between UAVs \\ 
\hline  
$\beta_0$ & LoS path loss at a reference distance \\
\hline  
$\gamma_{m,t}^{\textrm{A}}$ & SNR of signals received by passive UAV $m$ \\ 
\hline 
$\sigma^2$ & Variance of measurement error \\
\hline  
$E_{m,t}$ & Energy consumption of the active UAV\\ 
\hline  
$k_{m,t}$ & Distance between the BS and passive UAV $m$ \\
\hline  
$\boldsymbol{s}_{\textrm{B}}$ & Position of the BS\\ 
\hline 
$\chi_{m,t}$ & Elevation angle of passive UAV $m$ \\
\hline  
$L_{\textrm{FS}}$ & Free-space path loss \\ 
\hline 
$l_{m,t}^{\textrm{LoS}}$ & LoS path loss from UAV $m$ to the BS \\
\hline  
${\rm Pr}\left(l_{m,t}^{\textrm{LoS}}\right)$ & Probability of LoS  \\ 
\hline 
$l_{m,t}^{\textrm{NLoS}}$ & NLoS path loss from UAV $m$ to the BS \\
\hline  
$D$ & Data size of the distance information \\ 
\hline 
$\gamma_{m,t}^{\textrm{B}}$ & SNR of signals received at the BS \\
\hline  
$W$ & Bandwidth \\ 
\hline 
$\epsilon^{2}$ & Variance of Gaussian noise \\
\hline  
$T_{m,t}^{\textrm{A}}$ & Transmission delay between UAVs \\ 
\hline 
$\hat{\boldsymbol{r}}_{t}$ & Distance measurement information \\
\hline  
$\boldsymbol{r}_{t}$ & Actual distance \\ 
\hline 
$n_{m,t}$ & Measurement information error \\
\hline  
$T_{m,t}^{\textrm{B}}$ & Transmission delay from passive UAV $m$ to the BS \\ 
\hline
$V$ & Number of time slots \\
\hline  
$\hat{\boldsymbol{s}}_{t}$ & Estimated position of the target UAV\\
\hline
\end{tabular}}    
\vspace{-0.5cm}
\end{table}
\subsection{Transmission Model}
Here, we introduce the models for transmission links a) from the active UAV to the target UAV and then reflected to passive UAVs, b) from passive UAVs to the ground BS.
\subsubsection{Active UAV-Target UAV-Passive UAV Links} In our model, the active UAV transmits a signal $a_t$ to the target UAV. 
We assume that there is no occlusion in the path from the active UAV to the target UAV, and paths from the target UAV to passive UAVs. Let $\tau_{m,t}$ denote the time of transmitting signal $a_t$ from the active UAV to passive UAV $m$ via the target UAV. Then, $\tau_{m,t}$ can be given by
\begin{equation}\label{time}
    \tau_{m,t}=\frac{r_{m,t}\left(\boldsymbol{u}_{0,t},\boldsymbol{s}_t,\boldsymbol{u}_{m,t}\right)}{c},
\end{equation}where  $c$ is the speed of light and $r_{m,t}\left(\boldsymbol{u}_{0,t},\boldsymbol{s}_t,\boldsymbol{u}_{m,t}\right)=d_{0,t}\left(\boldsymbol{u}_{0,t},\boldsymbol{s}_t\right)+d_{m,t}\left(\boldsymbol{s}_{t},\boldsymbol{u}_{m,t}\right)$ is the distance from the active UAV to the target UAV and then from the target UAV to passive UAV $m$ with $d_{0,t}\left(\boldsymbol{u}_{0,t},\boldsymbol{s}_t\right)=\Vert \boldsymbol{u}_{0,t}-\boldsymbol{s}_{t} \Vert$ being the distance between the active UAV and the target UAV located at $\boldsymbol{s}_t=\left[x_t,y_t,z_t\right]^T$ and $d_{m,t}\left(\boldsymbol{s}_{t},\boldsymbol{u}_{m,t}\right)=\Vert \boldsymbol{s}_{t}-\boldsymbol{u}_{m,t} \Vert$ being the distance between the target UAV and passive UAV $m$. 

\textcolor{black}{Since less obstacles exist in the sky, we use a line-of-sight (LoS) transmission model for the links between the active UAV and passive UAVs \cite{UAV2,UAV3}.} Then,
the signals transmitted from the active UAV, reflected by the target UAV, and received by passive UAV $m$ at time slot $t$ is given by 
\begin{equation}
    y_{m,t}=\sqrt{p_{0,t}} h_{m,t} x_{m,t}h_{0,t}a_{t-\tau_{m,t}}+w_{m,t},
\end{equation}where $p_{0,t}$ is the transmit power of the active UAV at time slot $t$, $x_{m,t}$ represents the scattering coefficient of the target UAV \cite{1+1}, and $w_{m,t}$ is Gaussian noise with zero mean and $\epsilon^2$ variance. $h_{0,t}=\sqrt{\beta_0}d_{0,t}^{-1}\left(\boldsymbol{u}_{0,t},\boldsymbol{s}_{t}\right)$ represents the path loss from the active UAV to the target UAV, and $h_{m,t}=\sqrt{\beta_0}d_{m,t}^{-1}\left(\boldsymbol{u}_{m,t},\boldsymbol{s}_{t}\right)$ represents the path loss from the target UAV to passive UAV $m$ with $\sqrt{\beta_0}$ being the LoS path loss at a reference distance \cite{YZ}. We use LoS links to model the link between the active UAV and the target UAV and the links between the target UAV and passive UAVs.

At passive UAV $m$, the signal-to-noise ratio (SNR) of the signal transmitted by the active UAV and reflected by the target UAV is given by \cite{occlusion}
\begin{equation}\label{SNR_passiveUAV}
    \gamma_{m,t}^\textrm{A}\left(\boldsymbol{u}_{0,t},\boldsymbol{u}_{m,t},p_{0,t}\right)=\frac{p_{0,t}| h_{m,t}x_{m,t}h_{0,t}|^2}{\epsilon^2}.
\end{equation}

\textcolor{black}{From \eqref{SNR_passiveUAV}, we see that the SNR of each passive UAV depends on the transmit power of the active UAV and the distance between the active UAV and the passive UAV via the target UAV.} The transmission delay from the active UAV to the target UAV and from the target UAV to passive UAV $m$ is given by
\begin{equation}
    T_{m,t}^\textrm{A}\left(\boldsymbol{u}_{0,t},\boldsymbol{u}_{m,t},p_{0,t}\right)=\frac{D_\textrm{A}}{W\log_2\left(1+\gamma_{m,t}^\textrm{A}\left(\boldsymbol{u}_{m,t}\right)\right)},
\end{equation}where $D_\textrm{A}$ is the size of the transmitting signals and $W$ is the bandwidth. The energy consumption of the active UAV is given by
\begin{equation}
    E_{m,t}\left(\boldsymbol{u}_{0,t},\boldsymbol{u}_{m,t},p_{0,t}\right)=p_{0,t}T_{m,t}^\textrm{A}\left(\boldsymbol{u}_{0,t},\boldsymbol{u}_{m,t},p_{0,t}\right).
\end{equation}

Due to the limited energy of the active UAV, the transmit power of the active UAV must be optimized to minimize the positioning error of the target UAV while satisfying the energy consumption requirements of the active UAV.

\subsubsection{Passive UAV-BS Links}
Passive UAVs require to use their received signals to calculate the distance $\hat{r}_{m,t}$ from the active UAV to the target UAV and then from the target UAV to the passive UAV. Then, each passive UAV will transmit its calculated distance $\hat{r}_{m,t}$ to the BS. \textcolor{black}{Since the ground communications may interfere the transmission between UAVs and the BS,} we use probabilistic LoS and non-line-of sight (NLoS) links to model the links between passive UAVs and the BS. The LoS and NLoS path loss of passive UAV $m$ transmitting signals to the BS located at $\boldsymbol{s}_{\textrm{B}}$ at time slot $t$ is given by
\begin{equation}
\begin{split}
    l_{m,t}^{\rm LoS}&\left(\boldsymbol{u}_{m,t}\right) \\
    &=L_{\rm FS}\left(k_0\right)+10\mu_{\rm LoS}\log\left(k_{m,t}\left(\boldsymbol{u}_{m,t},\boldsymbol{s}_{\textrm{B}}\right)\right)+\lambda_{\sigma_{\rm LoS}},
    \end{split}
\end{equation}
\begin{equation}
\begin{split}
    l_{m,t}^{\rm NLoS}&\left(\boldsymbol{u}_{m,t}\right)=\\
    &L_{\rm FS}\left(k_0\right)+10\mu_{\rm NLoS}\log\left(k_{m,t}\left(\boldsymbol{u}_{m,t},\boldsymbol{s}_{\textrm{B}}\right)\right)+\lambda_{\sigma_{\rm NLoS}},
\end{split}
\end{equation}where $L_{\rm FS}\left(k_0\right)=20\log\left(k_0f_0^{\textrm B}4\pi/c\right)$ is the free-space path loss with $k_0$ being the free-space reference distance and $f_0^{\textrm B}$ being the carrier frequency. $k_{m,t}\left(\boldsymbol{u}_{m,t},\boldsymbol{s}_{\textrm{B}}\right)$ is the distance between passive UAV $m$ and the BS at time slot $t$. $\lambda_{\sigma_{\rm LoS}}$ and $\lambda_{\sigma_{\rm NLoS}}$ are the shadowing random variables, which are Gaussian variables in dB with zero mean and $\left(\sigma_{\rm LoS}^{\textrm B}\right)^2$, $\left(\sigma_{\rm NLoS}\right)^2$ dB variances. The probability of LoS is given by
\begin{equation}
    {\rm Pr}\left(l_{m,t}^{\rm LoS}\left(\boldsymbol{u}_{m,t}\right)\right)=\left(1+X\exp\left(-Y\left[\chi_{m,t}-X\right]\right)\right)^{-1},
\end{equation}where $X$ and $Y$ are constants which are related to the environment factors, and $\chi_{m,t}$ is the elevation angle of passive UAV $m$ at time slot $t$, which satisfies $\sin\left(\chi_{m,t}\right)=\frac{z_{m,t}}{k_{m,t}\left(\boldsymbol{u}_{m,t},\boldsymbol{s}_{\textrm{B}}\right)}$. Therefore, the path loss from passive UAV $m$ to the BS at time slot $t$ is given by
\begin{equation}
\begin{split}
    \bar{l}_{m,t}\left(\boldsymbol{u}_{m,t}\right)=&{\rm Pr}\left(l_{m,t}^{\rm LoS}\left(\boldsymbol{u}_{m,t}\right)\right)\times l_{m,t}^{\rm LoS}\left(\boldsymbol{u}_{m,t}\right)\\&+\left(1-{\rm Pr}\left(l_{m,t}^{\rm LoS}\left(\boldsymbol{u}_{m,t}\right)\right)\right)\times l_{m,t}^{\rm NLoS}\left(\boldsymbol{u}_{m,t}\right).
\end{split}
\end{equation}

\textcolor{black}{We assume that passive UAVs use an orthogonal frequency division multiple access (OFDMA) technique \cite{RVA}.} 
The SNR of the signal transmitted from passive UAV $m$ to the BS  at time slot $t$ is given by
\begin{equation}\label{SNR_BS}
    \gamma_{m,t}^\textrm{B}\left(\boldsymbol{u}_{m,t}\right)=\frac{p_{m,t}}{\epsilon^2}10^{-\bar{l}_{m,t}\left(\boldsymbol{u}_{m,t}\right)/10},
\end{equation}where $p_{m,t}$ is the transmit power of passive UAV $m$ at time slot $t$. \textcolor{black}{Hence, the SNR of the BS changes as the transmit powers of passive UAVs and the positions of passive UAVs vary.} The transmission delay from passive UAV $m$ to the BS at time slot $t$ is given by
\begin{equation}
    T_{m,t}^{\textrm{B}}\left(\boldsymbol{u}_{m,t}\right)=\frac{D_\textrm{B}}{W\log_2\left(1+\gamma_{m,t}^\textrm{B}\left(\boldsymbol{u}_{m,t}\right)\right)},
\end{equation}where $D_\textrm{B}$ is the data size of the distance information transmitted from passive UAVs to the BS.
\subsection{Model for Positioning}
Let $\hat{\boldsymbol{r}}_t=\left[\hat{r}_{1,t},\cdots,\hat{r}_{4,t}\right]^T$ be the distance measurement information received by the BS from passive UAVs. Then, the BS uses $\hat{\boldsymbol{r}}_t$ to estimate the position of the target UAV. A two-stage weighted least-squares (TSWLS) method \cite{TSWLS} is exploited to determine the position of the target UAV. Hence,
we assume that the distance measurements $\hat{\boldsymbol{r}}_{t}$ from the active UAV to passive UAV $m$ via the target UAV involves an error, and can be expressed by $\hat{r}_{m,t}=r_{m,t}+n_{m,t}\left(p_{0,t},\boldsymbol{u}_{0,t},\boldsymbol{u}_{m,t}\right)$, where $n_{m,t}\left(p_{0,t},\boldsymbol{u}_{0,t},\boldsymbol{u}_{m,t}\right)$  \textcolor{black}{represents the error between the measured distance $\hat{r}_{m,t}$ and the truth distance $r_{m,t}$} and is the independent Gaussian measurement error with zero mean and variance $\sigma_{m,t}^2\left(\boldsymbol{u}_{0,t},\boldsymbol{u}_{m,t},p_{0,t}\right)$ \cite{variance}.
Based on the distance measurement information $\hat{\boldsymbol{r}}_t$, 3D position of the controlled UAVs $\boldsymbol{U}_t=\left[\boldsymbol{u}_{0,t},\cdots,\boldsymbol{u}_{4,t}\right]^T$ and the transmit power $p_{0,t}$ of the active UAV at time slot $t$,
the estimated position of the target UAV $\hat{\boldsymbol{s}}_t\left(\boldsymbol{U}_{t},p_{0,t}\right)$ can be obtained via the TSWLS method in \cite{TSWLS}. 

\subsection{Problem Formulation}
Given the defined system model, our goal is to minimize the positioning error  \textcolor{black}{$\sum_{t=1}^{V} \sqrt{\left(\hat{\boldsymbol{s}}_{t}\left(\boldsymbol{U}_{t},p_{0,t}\right)-\boldsymbol{s}_{t}\right)^2}$} between the estimated position $\hat{\boldsymbol{s}}_t\left(\boldsymbol{U}_{t},p_{0,t}\right)$ and the actual position $\boldsymbol{s}_t$ of the target UAV over a time period $T$ that consists of $V$ time slots under the delay and movement constraints of UAVs, \textcolor{black}{where $\left(\hat{\boldsymbol{s}}_{t}\left(\boldsymbol{U}_{t},p_{0,t}\right)-\boldsymbol{s}_{t}\right)^2$ represents the square of the positioning error between the estimated position and the actual position of the target UAV at time slot $t$}. This minimization problem includes optimizing the transmit power of the active UAV and the trajectories of passive and active UAVs. The optimization problem is given by
\begin{equation}\label{optimization}
    \min_{p_{0,t},\boldsymbol{\varphi}_{t},\boldsymbol{\phi}_{t}} \sum_{t=1}^{V} \sqrt{\left(\hat{\boldsymbol{s}}_t\left(\boldsymbol{U}_{t},p_{0,t}\right)-\boldsymbol{s}_{t}\right)^2},\quad\quad\quad\quad\quad\quad\quad\quad
\end{equation}
\begin{flalign*}
    {\rm s.t.}\quad
    &E_{m,t}\leqslant E_{\textrm{max}}, \tag{\ref{optimization}a}\label{power}\\
    &T^{\textrm{B}}_{m,t}\left(\boldsymbol{u}_{m,t}\right)\leqslant \xi,\quad\forall{m\in\mathcal{M}}\tag{\ref{optimization}b}\label{T},\\
     &\varphi^{\textrm{min}}\leqslant \varphi_{m,t}\leqslant \varphi^{\textrm{max}}, \quad\forall{m\in\mathcal{M}}\tag{\ref{optimization}c}\label{theta},\\
     &\phi^{\textrm{min}}\leqslant \phi_{m,t}\leqslant \phi^{\textrm{max}}, \quad\forall{m\in\mathcal{M}}\tag{\ref{optimization}d}\label{phi},\\
     &L_{\textrm{min}}\leqslant \Vert\boldsymbol{u}_{m,t+1}-\boldsymbol{s}_{t+1}\Vert\leqslant L_{\textrm{max}}, \quad\forall{m\in\mathcal{M}} \tag{\ref{optimization}e}\label{TC},\\&L_{\textrm{min}}\leqslant \Vert\boldsymbol{u}_{m,t+1}-\boldsymbol{u}_{m',t+1}\Vert\leqslant L_{\textrm{max}}, \ \forall{m,m'\in\mathcal{M}}, \tag{\ref{optimization}f}\label{CC}
 \end{flalign*}
 where $p_{0,t}$ is the transmit power of the active UAV,
$\boldsymbol{\varphi}_{t}=\left[\varphi_{0,t},\ldots,\varphi_{4,t}\right]^T$ and $\boldsymbol{\phi}_t=\left[\phi_{0,t},\ldots,\phi_{4,t}\right]^T$ are the yaw angle vector and the pitch angle vector for the active UAV and passive UAVs, respectively. 
\eqref{power} is a maximum energy consumption constraint for the active UAV, \eqref{T} is the delay needed to transmit distance information from each passive UAV to the BS, \textcolor{black}{$E_{max}$ is the maximal energy of the active UAV, and $L_{max}$ is the maximal distance between any two UAVs to ensure the accurate UAV positioning.} \eqref{theta} and \eqref{phi} are the yaw angle and the pitch angle constraints for the controlled UAVs. \eqref{TC} is the constraint of the distance between a controlled UAV and the target UAV, and \eqref{CC} is the constraint of the distance between any two controlled UAVs.

The problem \eqref{optimization} is challenging to solve by conventional optimization algorithms due to the following reasons. First, since the Hessian matrix of objective function in \eqref{optimization} is not a positive semi-definite matrix, the problem \eqref{optimization} is non-convex. Second, the BS must know the coordinates of the target UAV to optimize the transmit power of the active UAV and trajectories of controlled UAVs using optimization methods. However, the target UAV is moving and hence the BS may not be able to obtain the real-time position of the target UAV. 
To solve the optimization problem \eqref{optimization}, we use a distributed RL algorithm which finds the probability distribution of the sum of future rewards to estimate the expected value of the sum of future rewards accurately. The proposed method enables the active UAV to determine its transmit power and each controlled UAV to determine its trajectory using its individual observation. Hence, using distributed RL, the BS and controlled UAVs can minimize the positioning error of the target UAV. 
\section{Proposed Z Function Decomposition based RL}
In this section, we introduce a ZD-RL method to solve the optimization problem in \eqref{optimization}. \textcolor{black}{Compared to standard RL algorithms \cite{VDN} such as deep Q-network (DQN) that uses a neural network to directly estimate the expected value of the sum of future rewards,  the ZD-RL method aims to find the probability distribution of the sum of future rewards and capture richer distribution information, thus improving the efficiency of optimizing the transmit power of the active UAV and trajectories of controlled UAVs.
} Hence, the ZD-RL method can improve the efficiency of optimizing the transmit power of the active UAV and trajectories of controlled UAVs.
Next, we first introduce the components of the ZD-RL method. Then, the process of using the ZD-RL method to find the global optimal transmit power for the active UAV and trajectories for controlled UAVs is explained.
\subsection{Components of the ZD-RL method}
The ZD-RL method consists of six components: a) agents, b) actions, c) states, d) rewards, e) individual Z function, f) global Z function, which are specified as follows:
\begin{itemize}
    \item \emph{Agents}: The agents that perform the ZD-RL method are the controlled UAVs. Each passive UAV must decide its yaw angle and pitch angle and the active UAV must decide its transmit power, yaw angle, and pitch angle  at each time slot. 
    \item \emph{State space}: A state of each agent is used to describe the local environment of each controlled UAV. In particular, a state of each passive UAV consists of its 3D coordinates and the distance measurements from the active UAV to the target UAV, and then from the target UAV to the passive UAV. Hence, a state of a passive UAV $m$ at time slot $t$ is $\boldsymbol{o}_{m,t}=\left[x_{m,t},y_{m,t},z_{m,t},\hat{r}_{m,t}\right]$. Since the active UAV cannot obtain the distance measurement, and the BS does not need the distance measurement of the active UAV to estimate the position of the target UAV, the state of the active UAV is $\boldsymbol{o}_{0,t}=\left[x_{0,t},y_{0,t},z_{0,t}\right]$. The states of all agents at time slot $t$ can be represented by a vector $\boldsymbol{o}_{t}=\left[\boldsymbol{o}_{0,t},\ldots,\boldsymbol{o}_{4,t}\right]$. 
    \item \emph{Actions}: The action of each passive UAV is the yaw angle and the pitch angle and the action of the active UAV is the transmit power, the yaw angle and the pitch angle. Hence, an action of passive UAV $m$ at time slot $t$ can be expressed as $\boldsymbol{a}_{m,t}=\left[\varphi_{m,t},\phi_{m,t}\right]$, and an action of the active UAV at time slot $t$ is $\boldsymbol{a}_{0,t}=\left[p_{0,t},\varphi_{0,t},\phi_{0,t}\right]$. The actions of all controlled UAVs at time slot $t$ is $\boldsymbol{a}_{t}=\left[\boldsymbol{a}_{0,t},\cdots,\boldsymbol{a}_{4,t}\right]$.

    \item \emph{Reward}: The reward of each controlled UAV captures the positioning accuracy of the target UAV resulting from a selected action. Given the global state $\boldsymbol{o}_t$ and the selected action $\boldsymbol{a}_{t}$, the reward of each controlled UAV at time slot $t$ is $R_{t}\left(\boldsymbol{o}_t,\boldsymbol{a}_t\right)=-\sqrt{\left(\hat{\boldsymbol{s}}_t\left(\boldsymbol{U}_{t},p_{0,t}\right)-\boldsymbol{s}_{t}\right)^2}$. Note that, $R_{t}\left(\boldsymbol{o}_t,\boldsymbol{a}_t\right)$ increases as the positioning error in \eqref{optimization} decreases, which implies that maximizing the reward of each controlled UAV can minimize the positioning error.
    \item \emph{Individual Z function}: 
    Z function is defined as the sum of future reward under a given state $\boldsymbol{o}_{m,t}$, a selection action $\boldsymbol{a}_{m,t}$, and a policy $\pi$, which can be expressed as $Z\left(\boldsymbol{o}_{m,t},\boldsymbol{a}_{m,t}\right)=\sum_{t=0}^{\infty}\gamma^tR\left(\boldsymbol{o}_{m,t},\boldsymbol{a}_{m,t}\right)$, where $\gamma$ is a discounted factor. Given the definition, our purpose is to estimate the probability distribution of $Z\left(\boldsymbol{o}_{m,t},\boldsymbol{a}_{m,t}\right)$. 
    This is different from 
    DQN \cite{VDN} that uses a neural network to estimate the sum of  expected future reward. In particular, the relationship between Q function and our defined Z function is expressed as 
    \begin{equation}
    \begin{split}
    Q\left(\boldsymbol{o}_{m,t},\boldsymbol{a}_{m,t}\right)&=\mathbb{E}_{\pi}\left[Z\left(\boldsymbol{o}_{m,t},\boldsymbol{a}_{m,t}\right)\right]\\&=\mathbb{E}_{\pi}\left[\sum_{t=0}^{\infty}\gamma^tR\left(\boldsymbol{o}_{m,t},\boldsymbol{a}_{m,t}\right)\right].
    \end{split}
    \end{equation}
    The advantage of estimating Z function instead of Q function is that Q function values estimated using the probability distribution of Z function are more accurate compared to Q function values directly estimated by DQN \cite{DYS}. 
    Hence, the ZD-RL method ensures the stability and effectiveness of model convergence \cite{IQN}. Next, we introduce the process of estimating the probability distribution of Z function.  
    First, we introduce the cumulative distribution function (CDF) of $Z\left(\boldsymbol{o}_{m,t},\boldsymbol{a}_{m,t}\right)$, which is given by 
    \begin{equation}\label{CDF}
        F\left(z\right)=\mathbb{P}\left(Z\left(\boldsymbol{o}_{m,t},\boldsymbol{a}_{m,t}\right)\leqslant z\right),
    \end{equation}    where $F\left(z\right)$ represents the probability that $Z\left(\boldsymbol{o}_{m,t},\boldsymbol{a}_{m,t}\right)$ is smaller than a value $z$. To estimate the probability distribution of $Z\left(\boldsymbol{o}_{m,t},\boldsymbol{a}_{m,t}\right)$, we use a DNN.
    The input of the DNN is the individual state $\boldsymbol{o}_{m,t}$, individual action $\boldsymbol{a}_{m,t}$ and a probability value $\varsigma_i$, and the output is a value of Z function, such as $\hat{Z}_{\boldsymbol{\omega}_m}\left(\boldsymbol{o}_{m,t},\boldsymbol{a}_{m,t},\varsigma_i\right)$, where $\boldsymbol{\omega}_m$ is the parameters of the DNN. The relationship between the input of DNN and its output can be expressed as
    \begin{equation}\label{quantile}
        \varsigma_i=\mathbb{P}\left(Z\left(\boldsymbol{o}_{m,t},\boldsymbol{a}_{m,t}\right)\leqslant \hat{Z}_{\boldsymbol{\omega}_m}\left(\boldsymbol{o}_{m,t},\boldsymbol{a}_{m,t},\varsigma_i\right)\right).
    \end{equation}
    From \eqref{quantile}, we can see that Z function is to find a value of $\hat{Z}_{\boldsymbol{\omega}_m}\left(\boldsymbol{o}_{m,t},\boldsymbol{a}_{m,t},\varsigma_i\right)$ such that $\mathbb{P}\left(Z\left(\boldsymbol{o}_{m,t},\boldsymbol{a}_{m,t}\right)\leqslant \hat{Z}_{\boldsymbol{\omega}_m}\left(\boldsymbol{o}_{m,t},\boldsymbol{a}_{m,t},\varsigma_i\right)\right)=\varsigma_i$. Given the relationship between $\varsigma_i$ and $\hat{Z}_{\boldsymbol{\omega}_m}\left(\boldsymbol{o}_{m,t},\boldsymbol{a}_{m,t},\boldsymbol{\varsigma}_i\right)$, the next step is to determine the value of $\varsigma_i$ such that we can use less DNN outputs to estimate the entire probability distribution of $Z\left(\boldsymbol{o}_{m,t},\boldsymbol{a}_{m,t}\right)$. To this end, we use a quantile vector  $\boldsymbol{\varsigma}=\left[\varsigma_1,\cdots,\varsigma_N\right]$ with $\varsigma_i=\frac{i}{N},i=1,\cdots,N$.
    \item \emph{Global Z function}: The global Z function $Z_{\textrm{T}}\left(\boldsymbol{o}_t,\boldsymbol{a}_t\right)$ is used to estimate the probability distribution of all controlled UAVs' achievable future rewards at each global state $\boldsymbol{o}_t$ and action $\boldsymbol{a}_t$. Similarly to individual Z functions, the probability distribution of the global Z function is approximated by a set of global Z function values with a quantile vector $\boldsymbol{\varsigma}$, and the approximated global Z function is represented by $\hat{Z}_{\textrm{T}}\left(\boldsymbol{o}_{t},\boldsymbol{a}_{t},\boldsymbol{\varsigma}\right)$. Based on the distributional individual-global-max principle \cite{DFAC}, the relationship between $\hat{Z}_{\textrm{T}}\left(\boldsymbol{o}_{t},\boldsymbol{a}_{t},\boldsymbol{\varsigma}\right)$ and $\hat{Z}_{\boldsymbol{\omega}_m}\left(\boldsymbol{o}_{m,t},\boldsymbol{a}_{m,t},\boldsymbol{\varsigma}\right)$ is given by
    \begin{equation}\label{global}
    \setlength{\abovedisplayskip}{0.5pt}
\setlength{\belowdisplayskip}{0.5pt}
    \begin{split} 
    \hat{Z}_{\textrm{T}}\left(\boldsymbol{o}_{t},\boldsymbol{a}_{t},\boldsymbol{\varsigma}\right)&
    =\sum_{m=0}^4M\left(\boldsymbol{o}_{m,t},\boldsymbol{a}_{m,t},\boldsymbol{\varsigma}\right)\\+\sum_{m=0}^4&\left(\hat{Z}_{\boldsymbol{\omega}_m}\left(\boldsymbol{o}_{m,t},\boldsymbol{a}_{m,t},\boldsymbol{\varsigma}\right)-M\left(\boldsymbol{o}_{m,t},\boldsymbol{a}_{m,t},\boldsymbol{\varsigma}\right)\right),
    \end{split}
    \end{equation}where $M\left(\boldsymbol{o}_{m,t},\boldsymbol{a}_{m,t},\boldsymbol{\varsigma}\right)$ is the approximated expected value of $\hat{Z}_{\boldsymbol{\omega}_m}\left(\boldsymbol{o}_{m,t},\boldsymbol{a}_{m,t},\boldsymbol{\varsigma}\right)$ and can be written as $M\left(\boldsymbol{o}_{m,t},\boldsymbol{a}_{m,t},\boldsymbol{\varsigma}\right)=\frac{1}{N}\sum_{i=1}^N \hat{Z}_{\boldsymbol{\omega}_m}\left(\boldsymbol{o}_{m,t},\boldsymbol{a}_{m,t},\varsigma_i\right)$.
\end{itemize}
\subsection{Training of the ZD-RL Method}
Here, we describe the entire training process of the ZD-RL method for optimizing the transmit power of the active UAV and trajectories of all controlled UAVs. 
In particular, we will first introduce the loss function of the ZD-RL method. Then, we introduce the training procedures. 
The total loss of the ZD-RL method is defined as the sum of the pair-wise loss for two values $\varsigma_i,\varsigma_j$ based on quantile Huber loss \cite{quantile}, where $\varsigma_i,\varsigma_j\in\boldsymbol{\varsigma}$. Compared to mean-square-error (MSE) loss and mean absolute error (MAE) used in traditional RL, the quantile Huber loss can reduce the sensitivity to abnormal samples that deviate from the normal range. 
The total loss is
\begin{equation}\label{loss}
\begin{split}&\mathfrak{L}_{\textrm{T}}\left(\boldsymbol{\omega}_0,\cdots,\boldsymbol{\omega}_4\right)\\&=\frac{1}{N}\sum_{t=1}^V\sum_{i=1}^N\sum_{j=1}^N \lvert \varsigma_i-\mathds{1}_{\left\{u\left(\boldsymbol{o}_{t},\boldsymbol{a}_{t},\varsigma_i,\varsigma_j\right)\textless0\right\}}\rvert\frac{G\left(u\left(\boldsymbol{o}_{t},\boldsymbol{a}_{t},\varsigma_i,\varsigma_j\right)\right)}{\eta},
    \end{split}
\end{equation}where $\mathds{1}_{\left\{x\right\}}=1$ when $x\textless0$ and $\mathds{1}_{\left\{x\right\}}=0$, otherwise. $u\left(\boldsymbol{o}_{t},\boldsymbol{a}_{t},\varsigma_i,\varsigma_j\right)=R_{t}\left(\boldsymbol{o}_t,\boldsymbol{a}_t\right)+\gamma \hat{Z}_{\textrm{T}}\left(\boldsymbol{o}_{t+1},\boldsymbol{a}_{t+1},\varsigma_j\right)-\hat{Z}_{\textrm{T}}\left(\boldsymbol{o}_t,\boldsymbol{a}_t,\varsigma_i\right)$ with $\boldsymbol{a}_{m,t+1}=\arg\max_{\boldsymbol{a}'_{m}} M\left(\boldsymbol{o}_{m,t+1},\boldsymbol{a}'_{m},\boldsymbol{\varsigma}\right)$ \cite{LYW}. $G\left(u\left(\boldsymbol{o}_{t},\boldsymbol{a}_{t},\varsigma_i,\varsigma_j\right)\right)$ is given by
\begin{equation*}
    \begin{split}G&\left(u\left(\boldsymbol{o}_{t},\boldsymbol{a}_{t},\varsigma_i,\varsigma_j\right)\right)\\&=\begin{cases}
    \frac{1}{2}\left(u\left(\boldsymbol{o}_{t},\boldsymbol{a}_{t},\varsigma_i,\varsigma_j\right)\right)^2, & \text{if}\quad \lvert u\left(\boldsymbol{o}_{t},\boldsymbol{a}_{t},\varsigma_i,\varsigma_j\right) \rvert \leqslant\eta,\\
    \eta\left(\lvert u\left(\boldsymbol{o}_{t},\boldsymbol{a}_{t},\varsigma_i,\varsigma_j\right)\rvert-\frac{1}{2}\eta\right), & \text{otherwise},
    \end{cases}
    \end{split}
\end{equation*}where $\eta$ is a hyper-parameter that determines the emphasis of Huber loss on MSE or MAE. \textcolor{black}{Here, using function $G\left(u\left(\boldsymbol{o}_{t},\boldsymbol{a}_{t},\varsigma_i,\varsigma_j\right)\right)$ can balance the sensitivity of MSE to large errors and the robustness of MAE to outliers and thus incorporating the strengths of both MSE and MAE. This is because the MSE loss function $\frac{1}{2}\left(u\left(\boldsymbol{o}_{t},\boldsymbol{a}_{t},\varsigma_i,\varsigma_j\right)\right)^2$ is highly sensitive to outliers since it squares the errors, which can destabilize learning in the presence of noise or anomalies.  The MAE loss function $\lvert u\left(\boldsymbol{o}_{t},\boldsymbol{a}_{t},\varsigma_i,\varsigma_j\right)\rvert$ is less sensitive to outliers when dealing with smaller errors. }

The training process consists of the following three steps: 

\newcommand{\tabincell}[2]{\begin{tabular}{@{}#1@{}}#2\end{tabular}}  
\begin{table}
\centering 
\small
 \begin{tabular}{l}
 \toprule  
 \textbf{Algorithm 1} ZD-RL Method for Solving Problem \eqref{optimization}\\
 \midrule  
 1: Initialize the DNN parameters $\boldsymbol{\omega}_m$ of each controlled UAV, \\ \quad a quantile vector $\boldsymbol{\varsigma}$.\\
 2: \textbf{for} each iteration \textbf{do}\\
 3. \quad\textbf{for} each controlled UAV $m$ \textbf{do}\\
 4. \quad\quad\textbf{for} each time slot $t$ \textbf{do}\\
 5. \quad\quad\quad Observe the observation $\boldsymbol{o}_{m,t}$.\\
 6: \quad\quad\quad  Select an action according to a $\epsilon$-greedy scheme.\\
 7: \quad\quad\quad  Calculate individual Z function values \\ \quad\quad\quad\quad $\hat{Z}_{\boldsymbol{\omega}_m}\left(\boldsymbol{o}_{m,t},\boldsymbol{a}_{m,t},\boldsymbol{\varsigma}\right)$ and $\hat{Z}_{\boldsymbol{\omega}_m}\left(\boldsymbol{o}_{m,t+1},\boldsymbol{a}_{m,t+1},\boldsymbol{\varsigma}\right)$. \\
 8: \quad\quad \textbf{end for}\\
 9: \quad\quad Controlled UAVs transmit $\boldsymbol{o}_{m,t}$, $\hat{Z}_{\boldsymbol{\omega}_m}\left(\boldsymbol{o}_{m,t},\boldsymbol{a}_{m,t},\boldsymbol{\varsigma}\right)$, \\ \quad\quad\quad\ and $\hat{Z}_{\boldsymbol{\omega}_m}\left(\boldsymbol{o}_{m,t+1},\boldsymbol{a}_{m,t+1},\boldsymbol{\varsigma}\right)$ to the BS.\\
 10: \quad \textbf{end for}\\
 11: \quad The BS calculates the reward and global Z function,\\\quad\quad\quad and transmits to controlled UAVs.\\
 12: \quad \textbf{for} each controlled UAV $m$ \textbf{do}\\
 13: \quad\quad Update $\boldsymbol{\omega}_{m}$ using $R\left(\boldsymbol{o}_t,\boldsymbol{a}_t\right)$, $\hat{Z}_{\textrm{T}}\left(\boldsymbol{o}_{t},\boldsymbol{a}_{t},\boldsymbol{\varsigma}\right)$ and \\\quad\quad\quad\ \  $\hat{Z}_{\textrm{T}}\left(\boldsymbol{o}_{t+1},\boldsymbol{a}_{t+1},\boldsymbol{\varsigma}\right)$ based on \eqref{update}.\\
 14: \quad \textbf{end for}\\
 15: \textbf{end for}\\
\bottomrule 
\end{tabular}
\vspace{-0.2cm}
\end{table}
\begin{itemize}
    \item \emph{Step 1 (training at controlled UAVs)}: Given a quantile vector $\boldsymbol{\varsigma}=\left[\varsigma_1,\cdots,\varsigma_N \right]$, each controlled UAV observes its local state $\boldsymbol{o}_{m,t}$, takes an action $\boldsymbol{a}_{m,t}$ according to a $\epsilon$-greedy algorithm,
    and calculates its individual Z function values $\hat{Z}_{\boldsymbol{\omega}_m}\left(\boldsymbol{o}_{m,t},\boldsymbol{a}_{m,t},\boldsymbol{\varsigma}\right)$, $\hat{Z}_{\boldsymbol{\omega}_m}\left(\boldsymbol{o}_{m,t+1},\boldsymbol{a}_{m,t+1},\boldsymbol{\varsigma}\right)$.
    Then, each UAV transmits its state $\boldsymbol{o}_{m,t}$, individual Z function values $\hat{Z}_{\boldsymbol{\omega}_m}\left(\boldsymbol{o}_{m,t},\boldsymbol{a}_{m,t},\boldsymbol{\varsigma}\right)$ and $\hat{Z}_{\boldsymbol{\omega}_m}\left(\boldsymbol{o}_{m,t+1},\boldsymbol{a}_{m,t+1},\boldsymbol{\varsigma}\right)$ to the BS.
    \item \emph{Step 2 (training at the BS)}: After collecting individual state and individual Z function values from all controlled UAVs, the BS calculates the reward $R_t\left(\boldsymbol{o}_t,\boldsymbol{a}_t\right)$ and the global Z function values $\hat{Z}_{\textrm{T}}\left(\boldsymbol{o}_{t},\boldsymbol{a}_{t},\boldsymbol{\varsigma}\right)$, $\hat{Z}_{\textrm{T}}\left(\boldsymbol{o}_{t+1},\boldsymbol{a}_{t+1},\boldsymbol{\varsigma}\right)$ based on \eqref{global}, and transmits $R_t\left(\boldsymbol{o}_t,\boldsymbol{a}_t\right)$, $\hat{Z}_{\textrm{T}}\left(\boldsymbol{o}_{t},\boldsymbol{a}_{t},\boldsymbol{\varsigma}\right)$, and $\hat{Z}_{\textrm{T}}\left(\boldsymbol{o}_{t+1},\boldsymbol{a}_{t+1},\boldsymbol{\varsigma}\right)$ to controlled UAVs. Here, the BS does not need to implement and update any neural networks.
    \item \emph{Step 3 (updating at controlled UAVs)}: Each UAV updates DNN parameters to approximate the probability distribution of its individual Z function using its collected global reward and global Z function values. The update of each controlled UAV $m$ is
    \begin{equation}\label{update}
        \begin{split}
        \boldsymbol{\omega}_{m}&=\boldsymbol{\omega}_{m}+\alpha_m\triangledown_{\boldsymbol{\omega}_{m}}\mathfrak{L}_{\textrm{T}}\left(\boldsymbol{\omega}_0,\cdots,\boldsymbol{\omega}_4\right),\\
        \end{split}
    \end{equation}where $\alpha_m$ is the step size. The entire training process of the ZD-RL method is summarized in Algorithm 1.
\end{itemize}
\subsection{Convergence, Implementation, and Complexity Analysis}
Next, we analyze the convergence, implementation and complexity of training the proposed ZD-RL method.

\emph{1) Convergence Analysis:} Here, we analyze the convergence of the proposed ZD-RL algorithm. We first analyze the gap between the optimal expected value of the individual Z function of controlled UAV $m$ and the expected value of individual Z function of controlled UAV $m$ obtained by the proposed ZD-RL method. Then, we show that this gap will converge to zero. In particular, the gap between the optimal expected value of individual Z function of controlled UAV $m$ and the expected value of individual Z function of controlled UAV $m$ obtained by the proposed ZD-RL method is
\begin{equation}\label{error}
    e\left(\boldsymbol{o}_{m,t},\boldsymbol{a}_{m,t}\right)=M\left(\boldsymbol{o}_{m,t},\boldsymbol{a}_{m,t}\right)-M^*\left(\boldsymbol{o}_{m,t},\boldsymbol{a}_{m,t}\right),
\end{equation}where $M^*\left(\boldsymbol{o}_{m,t},\boldsymbol{a}_{m,t}\right)=\mathbb{E}\left[Z^*\left(\boldsymbol{o}_{m,t},\boldsymbol{a}_{m,t}\right)\right]$ is the expected value of the optimal individual Z function of controlled UAV $m$ with respect to future Z functions (i.e., $Z^*\left(\boldsymbol{o}_{m,t+1},\boldsymbol{a}_{m,t+1}\right)$, $Z^*\left(\boldsymbol{o}_{m,t+2},\boldsymbol{a}_{m,t+2}\right),\cdots$).
From \eqref{error}, we can see that if the gap $e\left(\boldsymbol{o}_{m,t},\boldsymbol{a}_{m,t}\right)$ converges to zero, the proposed ZD-RL method converges \cite{pth}. To prove that the gap $e\left(\boldsymbol{o}_{m,t},\boldsymbol{a}_{m,t}\right)$ will finally converge to zero, we need to analyze how the gap changes as the number of training iterations increases. In particular, we define a distributional Bellman operator to find a relationship between the individual Z function of controlled UAV $m$ at two continuous time slots. 
In particular, the distributional Bellman operator of the individual Z function is defined as
 \begin{equation}\label{bellman}
     \mathcal{T}\left(Z\left(\boldsymbol{o}_{m,t},\boldsymbol{a}_{m,t}\right)\right)\overset{D}{:=}R\left(\boldsymbol{o}_{m,t},\boldsymbol{a}_{m,t}\right)+\gamma  Z\left(\boldsymbol{o}_{m,t+1},\boldsymbol{a}_{m,t+1}\right),
 \end{equation} where $\boldsymbol{a}_{m,t+1}=\arg\max_{\boldsymbol{a}'_{m}}  M\left(\boldsymbol{o}_{m,t+1},\boldsymbol{a}'_{m}\right)$.
Based on the above definition, the convergence of the proposed ZD-RL algorithm is shown in the following lemma. 

\noindent\textbf{Lemma 1.} The proposed ZD-RL method is guaranteed to converge to zero, if the following conditions are satisfied \cite{Q}:
\begin{itemize}
    \item[1)] The gap $e\left(\boldsymbol{o}_{m,t},\boldsymbol{a}_{m,t}\right)$ satisfies
    \begin{equation}
        \begin{split}
            e_{k+1}&\left(\boldsymbol{o}_{m,t},\boldsymbol{a}_{m,t}\right)\\&=\left(1-\alpha_m\right)e_k\left(\boldsymbol{o}_{m,t},\boldsymbol{a}_{m,t}\right)+\alpha_m F\left(\boldsymbol{o}_{m,t},\boldsymbol{a}_{m,t}\right),
        \end{split}
    \end{equation}where $F\left(\boldsymbol{o}_{m,t},\boldsymbol{a}_{m,t}\right)=R\left(\boldsymbol{o}_{m,t},\boldsymbol{a}_{m,t}\right)+\gamma M\left(\boldsymbol{o}_{m,t+1},\boldsymbol{a}_{m,t+1}\right)-M^*\left(\boldsymbol{o}_{m,t},\boldsymbol{a}_{m,t}\right)$.
    \item[2)] $\lvert\lvert\mathbb{E}\left[F\left(\boldsymbol{o}_{m,t},\boldsymbol{a}_{m,t}\right)\right]\rvert\rvert_{\infty}\leqslant\gamma\lvert\lvert e\left(\boldsymbol{o}_{m,t},\boldsymbol{a}_{m,t}\right)\rvert\rvert_{\infty},\forall \gamma\in\left(0,1\right)$, where $\lvert\lvert\cdot\rvert\rvert_{\infty}$ represents the infinite norm taking the maximum value of the absolute value of the elements, $\mathbb{E}\left[F\left(\boldsymbol{o}_{m,t},\boldsymbol{a}_{m,t}\right)\right]$ is the expected value of $F\left(\boldsymbol{o}_{m,t},\boldsymbol{a}_{m,t}\right)$ with respect to the state transition probability distribution.
    \item[3)] $\mathrm{Var}\left(\mathbb{E}\left[F\left(\boldsymbol{o}_{m,t},\boldsymbol{a}_{m,t}\right)\right]\right)\leqslant C_{\textrm{F}}\left(1+\lvert\lvert e\left(\boldsymbol{o}_{m,t},\boldsymbol{a}_{m,t}\right)\rvert\rvert^2_{\infty}\right)$, where $\mathrm{Var}\left(\mathbb{E}\left[F\left(\boldsymbol{o}_{m,t},\boldsymbol{a}_{m,t}\right)\right]\right)$ is the variance of $\mathbb{E}\left[F\left(\boldsymbol{o}_{m,t},\boldsymbol{a}_{m,t}\right)\right]$, and $C_{\textrm{F}}$ is a constant with $C_{\textrm{F}}\geqslant 0$.
\end{itemize}

\newenvironment{proof}{{\indent \indent \it Proof:}}{\hfill $\square$\par}
\begin{proof}
    See Appendix A.
\end{proof}

\begin{figure}[t]
 \setlength\abovecaptionskip{0.cm}
\setlength\belowcaptionskip{-0.25cm}
\centering
\includegraphics[width=8cm]{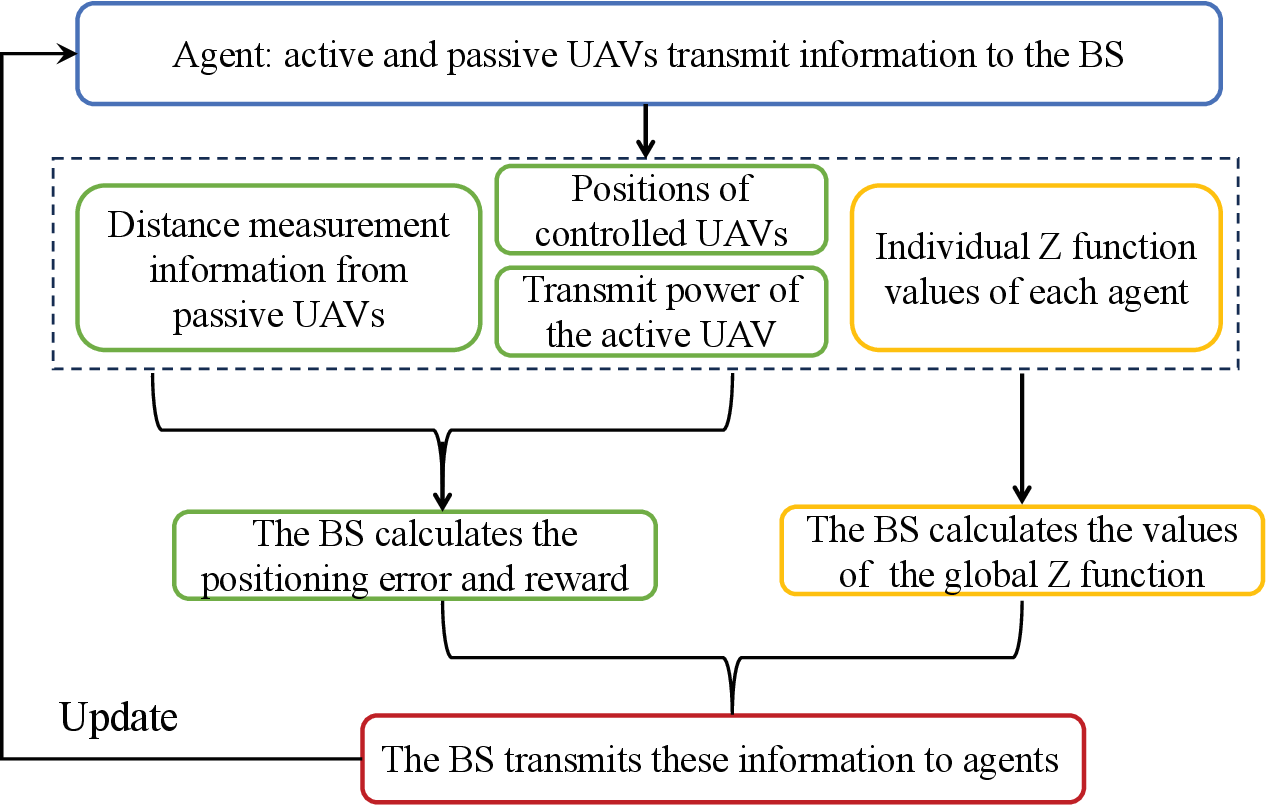}
\caption{\color{black}The flow chart of implementation.}
\label{implementation}
\vspace{-0.6cm}
\end{figure}

\emph{2) Implementation Analysis:} Next, we explain the implementation of the proposed ZD-RL method for UAV localization. 
The proposed ZD-RL method includes an offline training stage and an online decision-making stage. In the offline training phase, \textcolor{black}{as shown in Fig. \ref{implementation}}, each controlled UAV requires 1) the positioning error between the estimated position and the actual position of the target UAV and 2) the global Z function value to update its DNN parameters based on \eqref{loss} and \eqref{update}. 
To calculate the positioning error, the BS needs to collect the distance measurement information $\hat{r}_{m,t}$, the transmit power of the active UAV, and the positions of controlled UAVs. 
The distance information is estimated by the signals transmitted from the active UAV to the passive UAV and reflected by the target UAV. The transmit power of the active UAV is notified by the active UAV, and the positions of controlled UAVs are transmitted by controlled UAVs.
To calculate the global Z functions, the BS needs to collect individual Z functions as shown in \eqref{global} in our training stage. In the online decision-making stage, the well trained DNN can be directly used to determine the transmit power, yaw angle, and pitch angle of controlled UAVs. \textcolor{black}{From the implementation process, we see that the ZD-RL method enables each agent to train their deep neural networks parallelly and distributively. Hence, the designed ZD-RL method can be directly used in the scenario with more passive or active UAVs. In particular, when the number of agents increases, after all agents select and take actions, the BS will collect values of all individual Z functions from agents to calculate the global Z function values and collect positions and distance measurement information of all agents to calculate the positioning error of the target UAV. Thus, the ZD-RL method can adapt to the increase in the number of agents and enables the system to maintain its localization performance.}

\emph{3) Complexity Analysis:} The complexity of the proposed algorithm lies in training the DNN of each controlled UAV.
To analyze the complexity of training the designed ZD-RL method, we first assume that the value of the transmit power $p_{m,t}$ of controlled UAV $m$ at time slot $t$ is selected from a set of
$\left\{p^1_{m,t},\cdots,p^{N_{\textrm{P}}}_{m,t}\right\}$, the yaw angle $\varphi_{m,t}$ of controlled UAV $m$ is selected from a set $\left\{\varphi^1_{m,t},\cdots,\varphi^{N_1}_{m,t}\right\}$, and the pitch angle $\phi_{m,t}$ is selected from a set $\left\{\phi^1_{m,t},\cdots,\phi^{N_1}_{m,t}\right\}$ with $N_{\textrm{P}}$, $N_1$, and $N_2$ being the number of elements in their corresponding sets. Since we only consider optimizing the transmit power of the active UAV and the transmit power of passive UAVs are constant, we have $N_{\textrm{P}}=1$, when $m=1,\cdots,4$.
The interval of two yaw angles $\Delta \varphi_{m}$ is defined as $\Delta \varphi_{m}=\varphi^{i+1}_{m,t}-\varphi^{i}_{m,t},i=1,\cdots,N_1-1$ and the interval of two pitch angles $\Delta \phi_{m}$ is defined as $\Delta \phi_{m}=\phi^{i+1}_{m,t}-\phi^{i}_{m,t},i=1,\cdots,N_2-1$. Hence, the relationship between $N_1$, $N_2$ and the interval of angles $\Delta \varphi_{m}$ and $\Delta \phi_{m}$ is $N_1=\frac{\varphi^{N_1}_{m,t}-\varphi_{m,t}^1}{\Delta \varphi_{m}}+1,$ and $N_2=\frac{\phi^{N_2}_{m,t}-\phi_{m,t}^1}{\Delta \phi_{m}}+1$. 
Then, the complexity of training the designed ZD-RL method is shown in the following proposition. 

\noindent\textbf{Proposition 1.} The time complexity of training the proposed ZD-RL method is
\begin{equation}
    \begin{split}
    \mathcal{O}&\left(\sum_{l=1}^{L-1}l_il_{i+1}+\lvert\boldsymbol{o}_{m,t}\rvert l_1+N l_{L}\right.\\&\left.+l_L\left(N_{\textrm{P}}\left(\frac{\varphi^{N_1}_{m,t}-\varphi_{m,t}^1}{\Delta \varphi_{m}}+1\right)\left(\frac{\phi^{N_2}_{m,t}-\phi_{m,t}^1}{\Delta \phi_{m}}+1\right)\right)\right),
   \end{split}
\end{equation}
where $\lvert\boldsymbol{o}_{m,t}\rvert$ is the size of state space, $l_i$ is the number of neurons in hidden layer $i$, $L$ is the number of hidden layers, $N$ is the number of elements in the quantile vector. 

\newenvironment{proof2}{{\indent \indent \it Proof:}}{\hfill $\blacksquare$\par}
\begin{proof}
Based on \cite{wshAoI}, at each iteration, the time-complexity of training ZD-RL method is $\mathcal{O}\left(\sum_{l=1}^{L-1}l_il_{i+1}+\lvert\boldsymbol{o}_{m,t}\rvert l_1+N l_{L}+\lvert\boldsymbol{a}_{m,t}\rvert l_L\right)$, where $\lvert\boldsymbol{a}_{m,t}\rvert$ is the size of action space.
Since $\lvert\boldsymbol{a}_{m,t}\rvert$ depends on the interval $\Delta \varphi_{m}$ of two adjacent yaw angles and the interval $\Delta \phi_{m}$ of two adjacent pitch angles, $\lvert\boldsymbol{a}_{m,t}\rvert$ can be given by
\begin{equation}\label{numberandangel}
    \lvert\boldsymbol{a}_{m,t}\rvert=N_{\textrm{P}}\times\left(\frac{\varphi_{m,t}^{N_1}-\varphi_{m,t}^{1}}{\Delta \varphi_{m}}+1\right)\times\left(\frac{\phi_{m,t}^{N_2}-\phi_{m,t}^{1}}{\Delta \phi_{m}}+1\right),
\end{equation}where $N_{\textrm{P}}=1$ when $m=1,\cdots,4$. This is because we only consider optimizing the transmit power of the active UAV and the transmit power of passive UAVs are constant.
Based on \eqref{numberandangel}, the time-complexity of training the proposed ZD-RL method is
\begin{equation}
\begin{split}
    \mathcal{O}&\left(\sum_{l=1}^{L-1}l_il_{i+1}+\lvert\boldsymbol{o}_{m,t}\rvert l_1+N l_{L}\right.\\&\left.+l_L\left(N_{\textrm{P}}\left(\frac{\varphi_{m,t}^{N_1}-\varphi_{m,t}^{1}}{\Delta \varphi_{m}}+1\right)\left(\frac{\phi_{m,t}^{N_2}-\phi_{m,t}^{1}}{\Delta \phi_{m}}+1\right)\right)\right).\end{split}
\end{equation} This completes the proof. 
\end{proof}

From proposition 1, we see that as the interval $\Delta \varphi_{m}$ and $\Delta \phi_{m}$ of two adjacent angles decreases, the time-complexity of training the proposed ZD-RL method at each iteration increases and hence the number of iterations that the ZD-RL method required to converge increases. 
However, when the intervals $\Delta \varphi_{m}$ and $\Delta \phi_{m}$ increases, the controlled UAVs may find better yaw angles and pitch angles for the target UAV localization thus improving localization performance.

\section{Controlled UAV Deployment for Target UAV Localization}
In this section, we aim to find the positions of contrlled UAVs that can minimum the positioning error of the target UAV. 
At each time slot, the relationship between the positions of controlled UAVs and the distance $r_{m,t}$ from the active UAV to the target UAV and then from the target UAV to passive UAV $m$ is given by
\begin{equation}\label{m1}
r_{m,t}=
d_{m,t}\left(\boldsymbol{u}_{m,t},\boldsymbol{s}_{t}\right)+d_{0,t}\left(\boldsymbol{u}_{0,t},\boldsymbol{s}_{t}\right),
\end{equation}

Taking differentiation at both sides of \eqref{m1}, we have
 \begin{equation}\label{m2}
 \begin{split}
     \textrm{d}r_{m,t}=&\left(\frac{x_t-x_{m,t}}{d_{m,t}}+\frac{x_t-x_{0,t}}{d_{0,t}}\right)\textrm{d}x_t\\&+\left(\frac{y_t-y_{m,t}}{d_{m,t}}+\frac{y_t-y_{0,t}}{d_{0,t}}\right)\textrm{d}y_t\\&+\left(\frac{z_t-z_{m,t}}{d_{m,t}}+\frac{z_t-z_{0,t}}{d_{0,t}}\right)\textrm{d}z_t, \quad m=1,2,3,4.
 \end{split}
 \end{equation}
 Then, we can rewrite \eqref{m2} as
 \begin{equation}\label{m5}
     \textrm{d}\boldsymbol{r}_t=\boldsymbol{M} \textrm{d}\boldsymbol{s}_{t} 
 \end{equation}where $\textrm{d}\boldsymbol{r}_t=\left[\textrm{d} r_{1,t},\textrm{d}r_{2,t},\textrm{d}r_{3,t},\textrm{d}r_{4,t}\right]^T$, $\textrm{d}\boldsymbol{s}_t=\left[\textrm{d} x_{t},\textrm{d}y_{t},\textrm{d}z_{t}\right]^T$,
 and 
 \begin{equation}\label{m3}
 \begin{split}
 &\boldsymbol{M}=\\&\begin{bmatrix}
         \frac{x_t-x_{1,t}}{d_{1,t}}+\frac{x_t-x_{0,t}}{d_{0,t}}&\frac{y_t-y_{1,t}}{d_{1,t}}+\frac{y_t-y_{0,t}}{d_{0,t}}&\frac{z_t-z_{1,t}}{d_{1,t}}+\frac{z_t-z_{0,t}}{d_{0,t}}\\
         \frac{x_t-x_{2,t}}{d_{2,t}}+\frac{x_t-x_{0,t}}{d_{0,t}}&\frac{y_t-y_{2,t}}{d_{2,t}}+\frac{y_t-y_{0,t}}{d_{0,t}}&\frac{z_t-z_{2,t}}{d_{2,t}}+\frac{z_t-z_{0,t}}{d_{0,t}}\\
         \frac{x_t-x_{3,t}}{d_{3,t}}+\frac{x_t-x_{0,t}}{d_{0,t}}&\frac{y_t-y_{3,t}}{d_{3,t}}+\frac{y_t-y_{0,t}}{d_{0,t}}&\frac{z_t-z_{3,t}}{d_{3,t}}+\frac{z_t-z_{0,t}}{d_{0,t}}\\
         \frac{x_t-x_{4,t}}{d_{4,t}}+\frac{x_t-x_{0,t}}{d_{0,t}}&\frac{y_t-y_{4,t}}{d_{4,t}}+\frac{y_t-y_{0,t}}{d_{0,t}}&\frac{z_t-z_{4,t}}{d_{4,t}}+\frac{z_t-z_{0,t}}{d_{0,t}}\\
     \end{bmatrix}.\end{split}
 \end{equation}

Based on \eqref{m5}, the positioning error between the estimated position $\hat{\boldsymbol{s}}_t$ and the actual position $\boldsymbol{s}_t$ of the target UAV in \eqref{optimization} at time slot $t$ can be expressed as $e_{t}=\sqrt{\left(\textrm{d}x_t\right)^2+\left(\textrm{d}y_t\right)^2+\left(\textrm{d}z_t\right)^2}$ \cite{e_t}. Hence, we have $e_{t}=\sqrt{\textrm{tr}\left(\mathbb{E}\left[\textrm{d}\boldsymbol{s}_{t}\textrm{d}\boldsymbol{s}_{t}^{T}\right]\right)}$, where $\textrm{tr}\left(\cdot\right)$ is the trace of the matrix. Then, the minimum value of the positioning error $e_t$ of the target UAV is shown in the following proposition. 

\noindent\textbf{Theorem 2.} If the distances between passive UAVs and the target UAV satisfy $d_{1,t}=d_{2,t}=d_{3,4}=d_{4,t}$, the minimum positioning error of the target UAV $e_t$ is 
\begin{equation}
    e_t=\sqrt{4k\left(L_{\textrm{min}}\right)^{2}\textrm{tr}\left(\left(\boldsymbol{M}^T\boldsymbol{M}\right)^{-1}\right)}.
\end{equation}

\begin{proof}
    See Appendix B.
\end{proof}

From Theorem 2, we can see that the minimum positioning error of the target UAV depends on the safety distance $L_{\textrm{min}}$ between any two UAVs in constraint \eqref{TC}, and the value of $\textrm{tr}\left(\left(\boldsymbol{M}^T\boldsymbol{M}\right)^{-1}\right)$  which relies on the positions of controlled UAVs. Theorem 2 also shows that as the distance between each controlled UAV and the target UAV is minimum (i.e., $d_{1,t}=d_{2,t}=d_{3,t}=d_{4,t}=L_{\textrm{min}}$), the positioning error can be minimized.

Based on Theorem 2, next, we can also derive the minimum positioning error of the target UAV when the position of the active UAV is given, which is shown in the following proposition.

\noindent\textbf{Lemma 2.} Given the positions of the target UAV $\boldsymbol{s}_t$ and the active UAV $\boldsymbol{u}_{0,t}$, if the distances from passive UAVs to the target UAV satisfy $d_{1,t}=d_{2,t}=d_{3,t}=d_{4,t}$, the minimum positioning error of the target UAV is 
\begin{equation}
    e_t=\frac{3}{2}\left(L_{\textrm{min}}+d_{0,t}\right)\sqrt{k},
\end{equation}where $k$ is a coefficient \cite{variance}.

\begin{proof}
    See Appendix C.
\end{proof}

From Lemma 2, we see that when the positions of the active UAV and the target UAV are given, the minimum positioning error only depends on the distance $L_{\textrm{min}}$ between each passive UAV and the target UAV.
\section{Simulation Results and Analysis}
\begin{table}[t!]
\caption{\label{parameters}Parameters}  
\centering 
\setlength{\abovecaptionskip}{0.cm}
\setlength{\tabcolsep}{0.8mm}{
\begin{tabular}{|c|c|c|c|}  
\hline  
$\mathbf{Parameters}$ & $\mathbf{Values}$ & $\mathbf{Parameters}$ & $\mathbf{Values}$ \\ 
\hline  
$c$ & 3$e^8$ m/s & $p_{m,t}$ & 5 W \\ 
\hline 
$\epsilon^2$ & -95 dBm & $W$ & 1 MHz \\ 
\hline  
$\left(\sigma_{\textrm{LoS}}^B\right)^2$ & 8.41 & $\left(\sigma_{\textrm{NLoS}}^B\right)^2$ & 33.78 \\ 
\hline  
$E_{\textrm{max}}$ & 100 kJ & $\xi$ & 1 s  \\ 
\hline  
$L_{\textrm{min}}$ & 100 m & $L_{\textrm{max}}$ & 10 km  \\ 
\hline  
$\phi_{\textrm{min}}$ & $-15^{o}$ & $\phi_{\textrm{max}}$ & $15^{o}$  \\ 
\hline  
$\varphi_{\textrm{min}}$ & $-15^{o}$ & $\varphi_{\textrm{max}}$ & $15^{o}$ \\ 
\hline 
 $D_{\textrm{B}}$ & 5 bit  & $V$ & 30 \\
\hline
$\mu_{\textrm{LoS}}^B$ & 2 & $\mu_{\textrm{NLoS}}^B$ & 2.4 \\
\hline
$Y$ & 0.13& $X$ & 11.9\\
\hline
\end{tabular}}    
\vspace{-0.15cm}
\end{table}

\begin{table}[t!]
\caption{\label{hyper}\textcolor{black}{Hyperparameters}} 
\centering 
\textcolor{black}{
\setlength{\abovecaptionskip}{0.cm}
\setlength{\tabcolsep}{0.8mm}{
\begin{tabular}{|c|c|}  
\hline  
$\mathbf{Hyperparameters}$ & $\mathbf{Values}$ \\ 
\hline  
Discounted factor $\gamma$ & 0.9 \\
\hline 
The number of hidden layers of each agent & 2 \\ 
\hline  
The number of neurons of each hidden layer & 64  \\ 
\hline  
Learning rate & 0.0005 \\ 
\hline  
The size of a batch & 512 \\ 
\hline  
The number of episodes of the target network per update & 200 \\ 
\hline  
The size of the replay buffer & 2000 \\ 
\hline  
\end{tabular}} }  
\vspace{-0.5cm}
\end{table}

For our simulations, five controlled UAVs and a BS jointly localize a target UAV. The moving speed of each controlled UAV is $v_{m,t}=10$ m/s and the time duration of a time slot is $\Delta_t=1$ s. We use the TSWLS method to estimate the position of the target UAV at each time slot \cite{TSWLS}. Other system parameters are listed in Table \ref{parameters} \textcolor{black}{and the training hyperparameters are listed in Table \ref{hyper}}. For comparison, we consider five baselines: a) independent DRL method in which each controlled UAV uses a DQN to optimize its trajectory without considering other controlled UAVs' movements and b) VD-RL method in which controlled UAVs collaboratively determine their trajectories to minimize positioning errors by summing individual Q function values to approximate the global Q function value \cite{VDN}.


\begin{table}[t!]
\caption{\label{complexity}\textcolor{black}{Training Complexity} }
\centering 
\textcolor{black}{
\setlength{\abovecaptionskip}{0.cm}
\setlength{\tabcolsep}{0.8mm}{
\begin{tabular}{|c|c|c|}  
\hline  
$\mathbf{Methods}$ & $\mathbf{Time\ per\ iteration  (s)}$ &  $\mathbf{Iterations}$ \\ 
\hline   
ZD-RL & 0.0090 & 180800\\ 
\hline  
VD-RL & 0.0083 & 216200\\ 
\hline  
Qtran & 0.0079 & 218200  \\ 
\hline  
Independent DRL & 0.0081& 224200  \\ 
\hline  
Mappo & 0.0147 & 301800\\ 
\hline  
\end{tabular}}}
\vspace{-0.5cm}
\end{table}

\begin{figure*}[h! ]
 \setlength\abovecaptionskip{0.cm}
\setlength\belowcaptionskip{-0.2cm}
\centering
\subfigure[]{
\includegraphics[width= 0.26\linewidth,height= 0.26\linewidth]{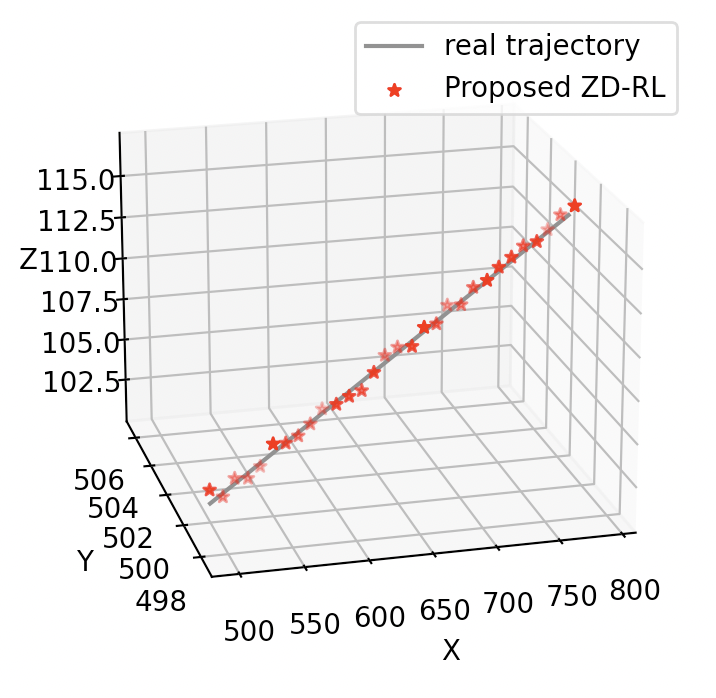}
}
\subfigure[]{
\includegraphics[width=  0.26\linewidth,height=  0.26\linewidth]{ 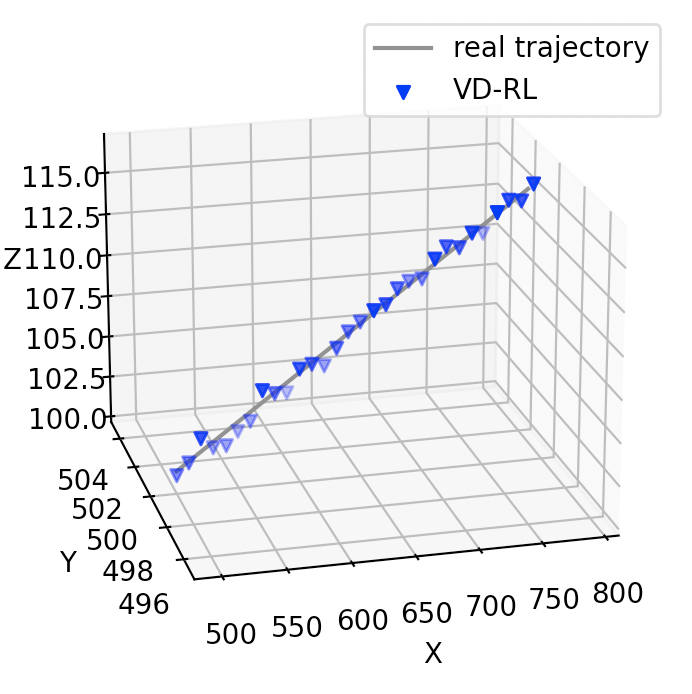}
}
\subfigure[]{
\includegraphics[width=  0.26\linewidth,height=  0.26\linewidth]{ 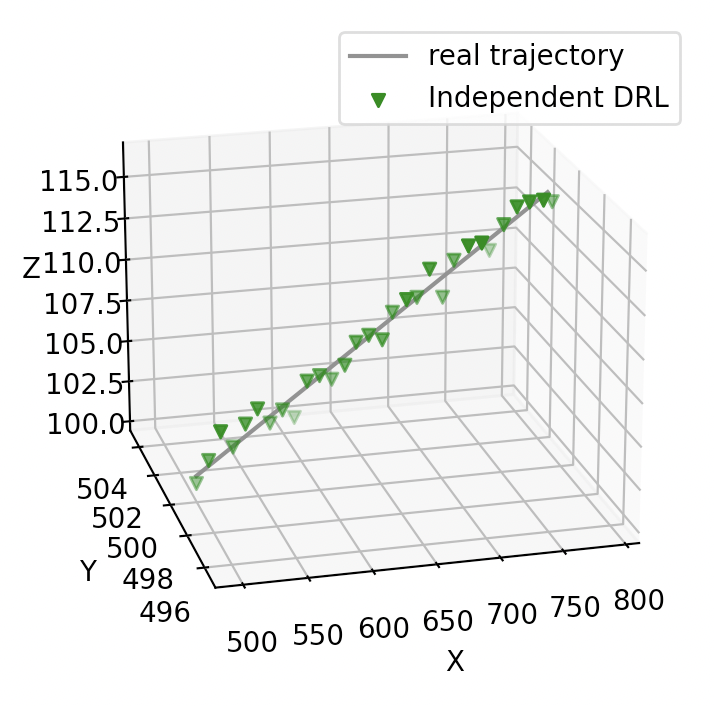}
}
\subfigure[]{
\includegraphics[width=  0.26\linewidth,height=  0.26\linewidth]{ 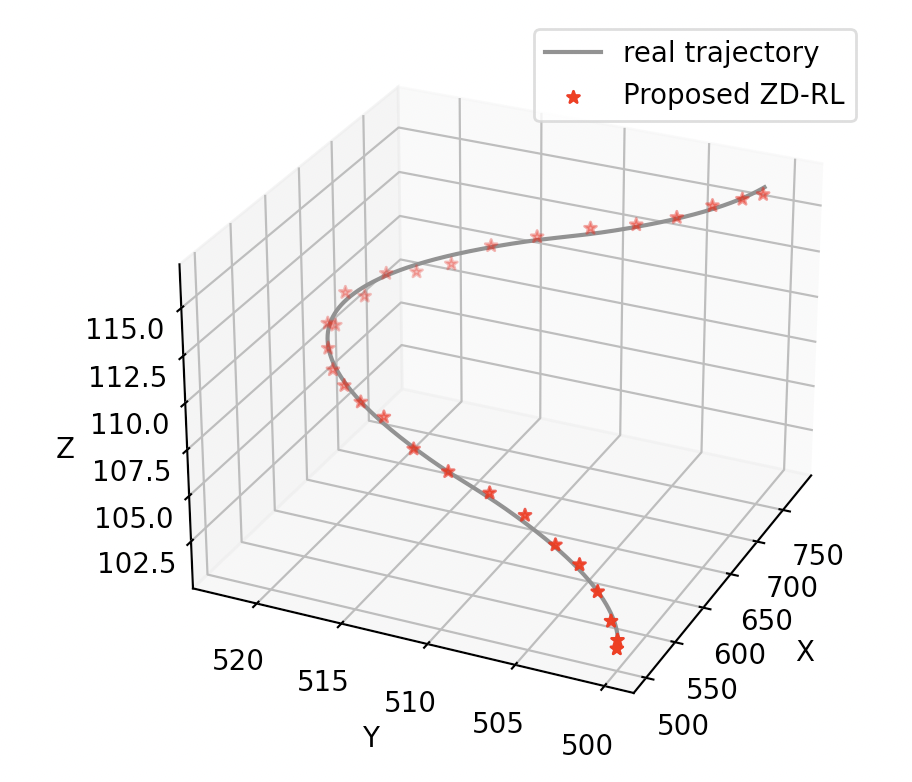}
}
\subfigure[]{
\includegraphics[width=  0.26\linewidth,height=  0.26\linewidth]{ 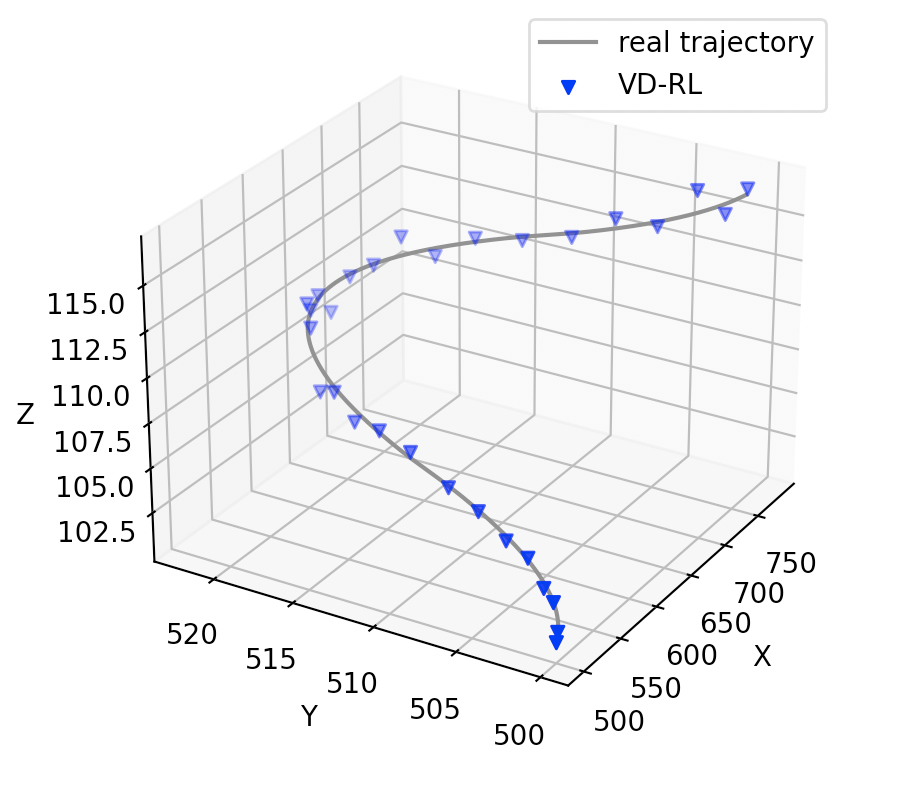}
}
\subfigure[]{
\includegraphics[width=  0.26\linewidth,height=  0.26\linewidth]{ 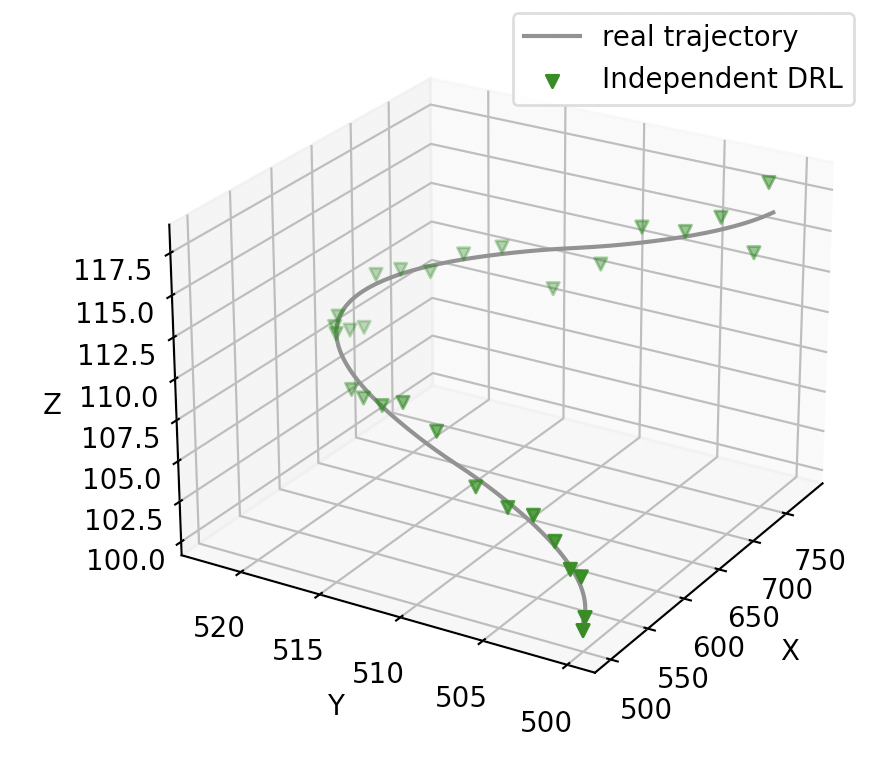}
}
\subfigure[]{
\includegraphics[width=  0.26\linewidth,height=  0.26\linewidth]{ 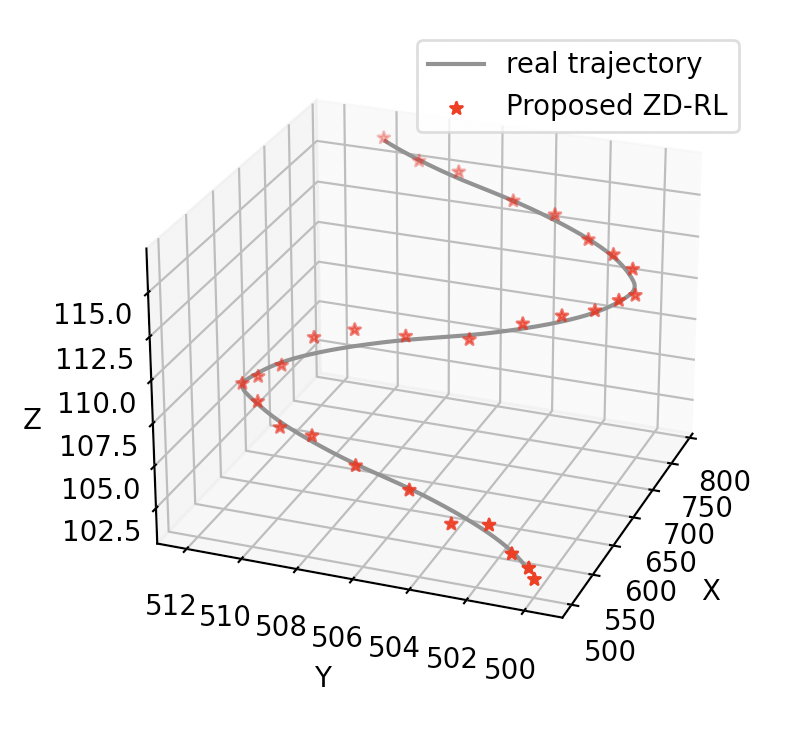}
}
\subfigure[]{
\includegraphics[width=  0.26\linewidth,height=  0.26\linewidth]{ 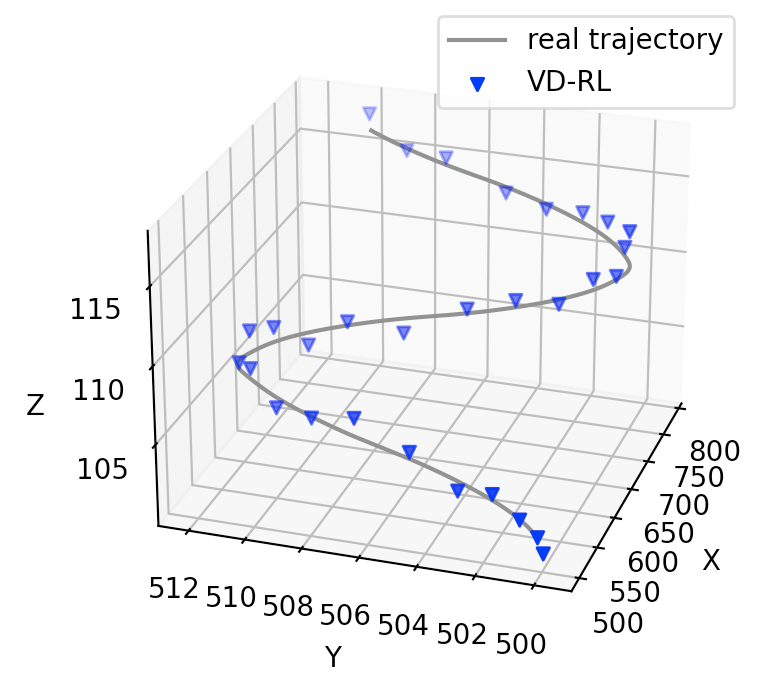}
}
\subfigure[]{
\includegraphics[width=  0.26\linewidth,height=  0.26\linewidth]{ 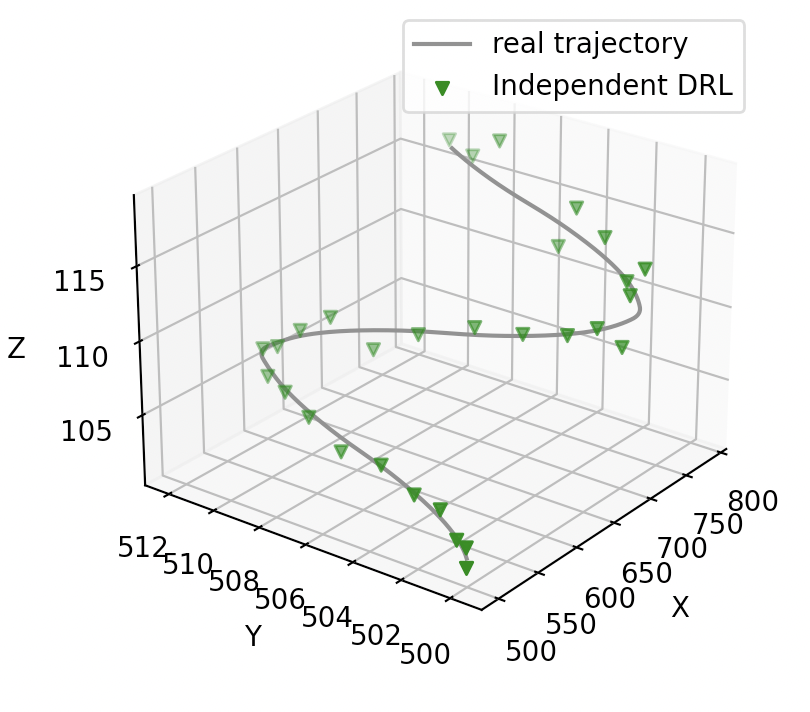}
}
\subfigure[]{
\includegraphics[width=  0.26\linewidth,height=  0.26\linewidth]{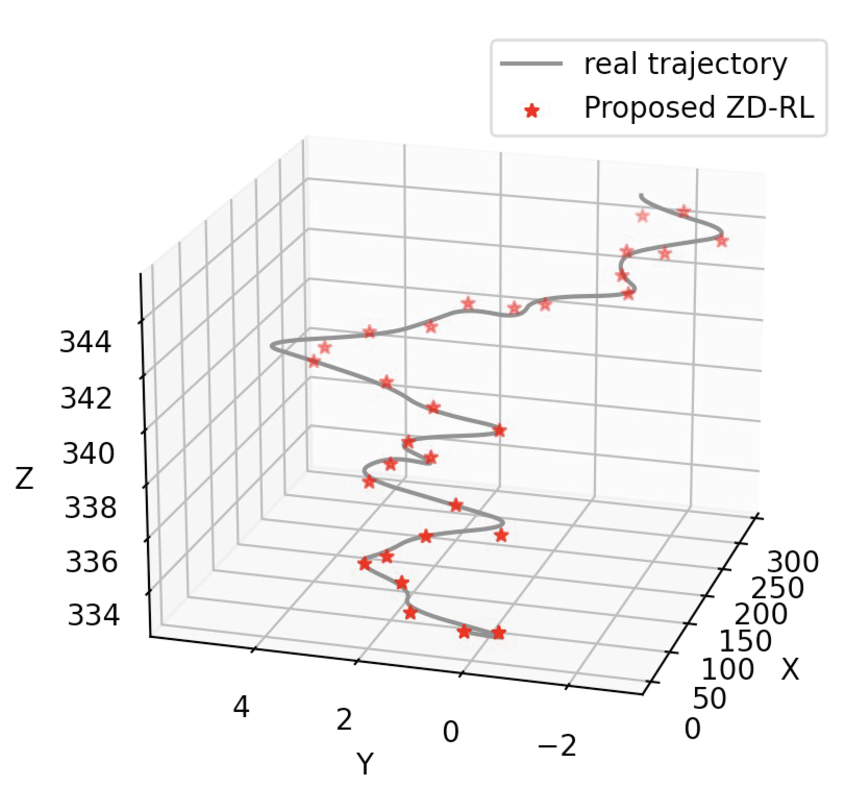}
}
\subfigure[]{
\includegraphics[width=  0.26\linewidth,height=  0.26\linewidth]{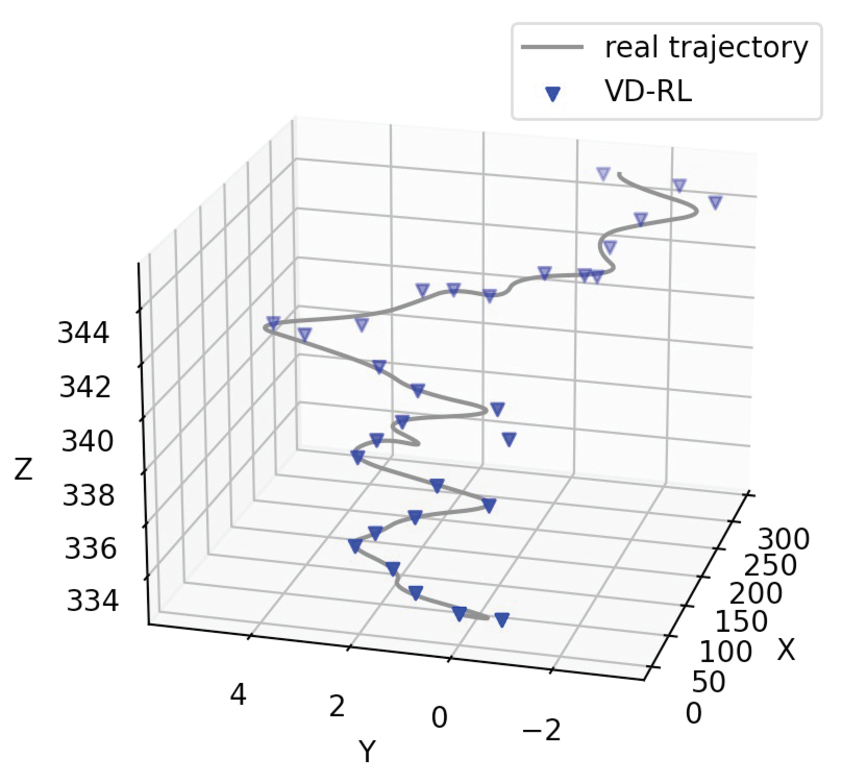}
}
\subfigure[]{
\includegraphics[width=  0.26\linewidth,height=  0.26\linewidth]{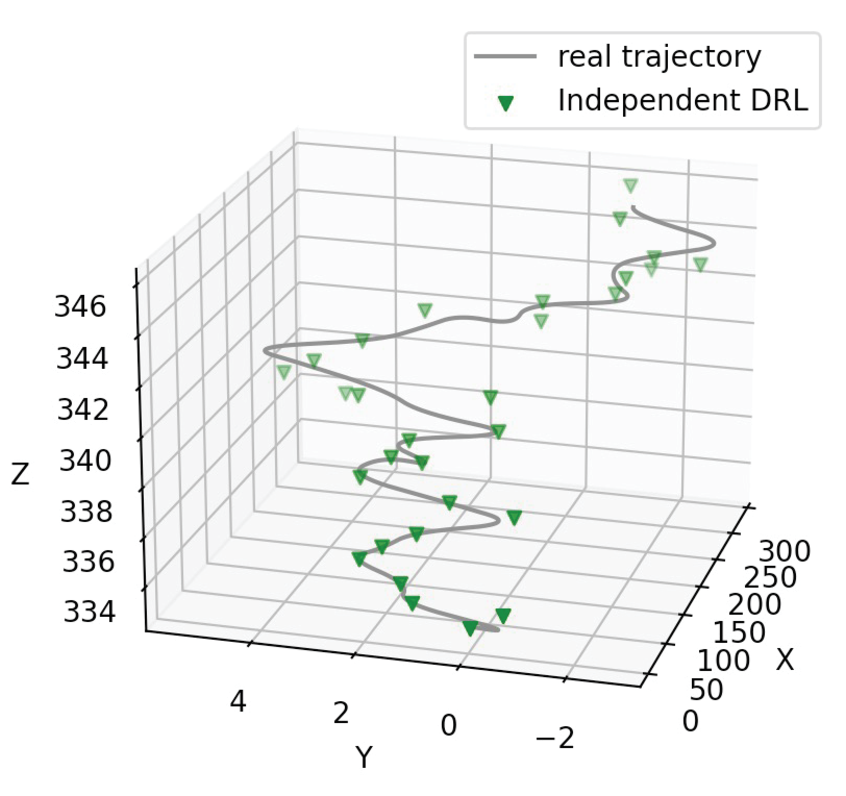}
}
\caption{\textcolor{black}{The actual trajectories of the target UAV and the estimated trajectories obtained by different methods.}}
\label{trajectory}
\vspace{-0.6cm}
\end{figure*}
Fig. \ref{trajectory} shows the actual and the estimated trajectories of the target UAV obtained by the considered algorithms. 
In Figs. \ref{trajectory}(a), \ref{trajectory}(b), and \ref{trajectory}(c), the target UAV moves in a straight line from the stating position (500 m, 500 m, 
100 m) to (789 m, 500 m, 
116 m) and five controlled UAVs are randomly distributed in a sphere of radius 1000 m centered on the target UAV. In Figs. \ref{trajectory}(d), \ref{trajectory}(e), and \ref{trajectory}(f), the target UAV moves in the curve of ``C''. In Figs. \ref{trajectory}(g), \ref{trajectory}(h), and \ref{trajectory}(i), the target UAV follows the curve of ``S''. \textcolor{black}{In Figs. \ref{trajectory}(j), \ref{trajectory}(k), and \ref{trajectory}(l), the real trajectory of the target UAV is generated by its movement from the starting position (0 m, 0 m, 333 m) and the target UAV selects the pitch angle and yaw angle randomly at each time slot.}  From Fig. \ref{trajectory}, we can also see that the gaps between the real trajectories and estimated trajectories obtained by the proposed ZD-RL increase as the trajectories of the target UAV become more complex. This is because as the trajectories of the target UAV becomes more complex, it becomes more difficult for the proposed ZD-RL method to control the trajectories of controlled UAVs to keep small distances with the target UAV in real time. 
From Fig. \ref{trajectory}, we can also see that the proposed method can estimate the target UAV position more accurately compared to the VD-RL, and independent DRL method. 
As the target UAV moves from the initial position to the end position, the gap between the actual positions and the positions estimated by the proposed ZD-RL method is small while the gap resulting from each baseline increases. This is due to the fact that, the proposed ZD-RL method enables controlled UAVs to cooperatively select the pitch angle and yaw angle based on the global Z function, which is generated by the BS using a set of individual Z functions thus the proposed ZD-RL method can accurately optimize the trajectories of controlled UAVs in time to track the target UAV as the target UAV moves in different trajectories. 


\begin{figure}[t!]
 \setlength\abovecaptionskip{0.cm}
\setlength\belowcaptionskip{-0.2cm}
\centering 
\includegraphics[width=8cm]{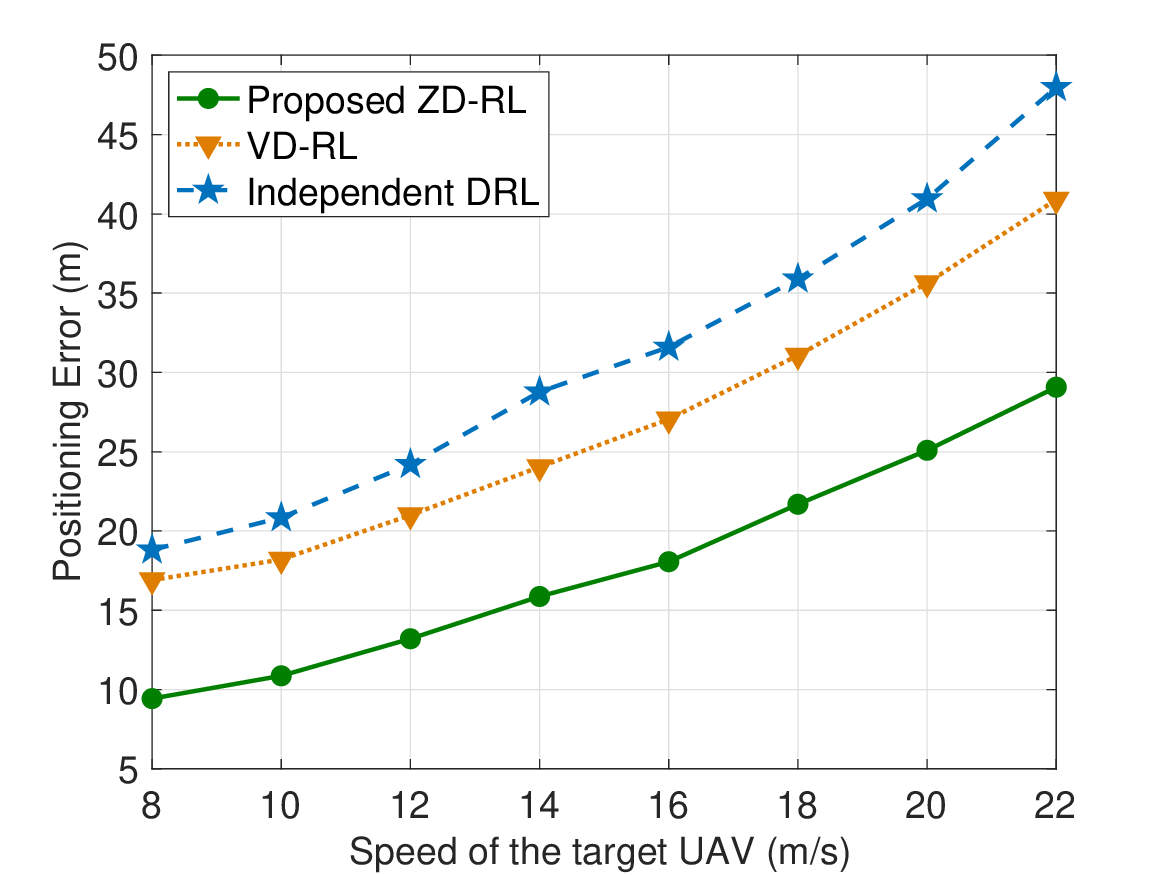}
\caption{Value of the positioning error as the speed of the target UAV varies.}
\label{speed} 
\vspace{-0.3cm}
\end{figure} 

Fig. \ref{speed} shows how the positioning error changes as the speed of the target UAV varies when the target UAV moves in the curve of ``S'' . In Fig. \ref{speed}, we can see that as the speed of the target UAV increases, the positioning errors of the considered algorithms increase. This is due to the fact that as the speed of the target UAV increases, controlled UAVs cannot follow the target UAV and the distances between the target UAV and controlled UAVs increase. 
Fig. \ref{speed} also shows that the proposed ZD-RL method can achieve up to 28.9\% and 39.6\% gains in terms of the positioning accuracy compared to the VD-RL method and independent DRL method, respectively, in the case that the target UAV moving at the speed of 22 m/s. The 28.9\% gain stems from the fact that the VD-RL method obtains the global value function by linearly calculating the sum of the expected value of future rewards at each controlled UAV. However, the proposed ZD-RL method calculates the global Z function using a set of global Z functions, which contains more interaction information with the environment thus being able to select pitch angle and yaw angle for controlled UAVs and optimize the transmit power for the target UAV to localize the target UAV accurately. The 39.6\% gain is because the proposed ZD-RL uses the global observation information and global reward generated by the BS to train DNN parameters of each controlled UAV and enables controlled UAVs to select accurate actions by learning the movements from each other thus improving the localization accuracy cooperatively. 

\begin{figure}[t!]
 \setlength\abovecaptionskip{0.cm}
\setlength\belowcaptionskip{-0.2cm}
\centering 
\includegraphics[width=8cm]{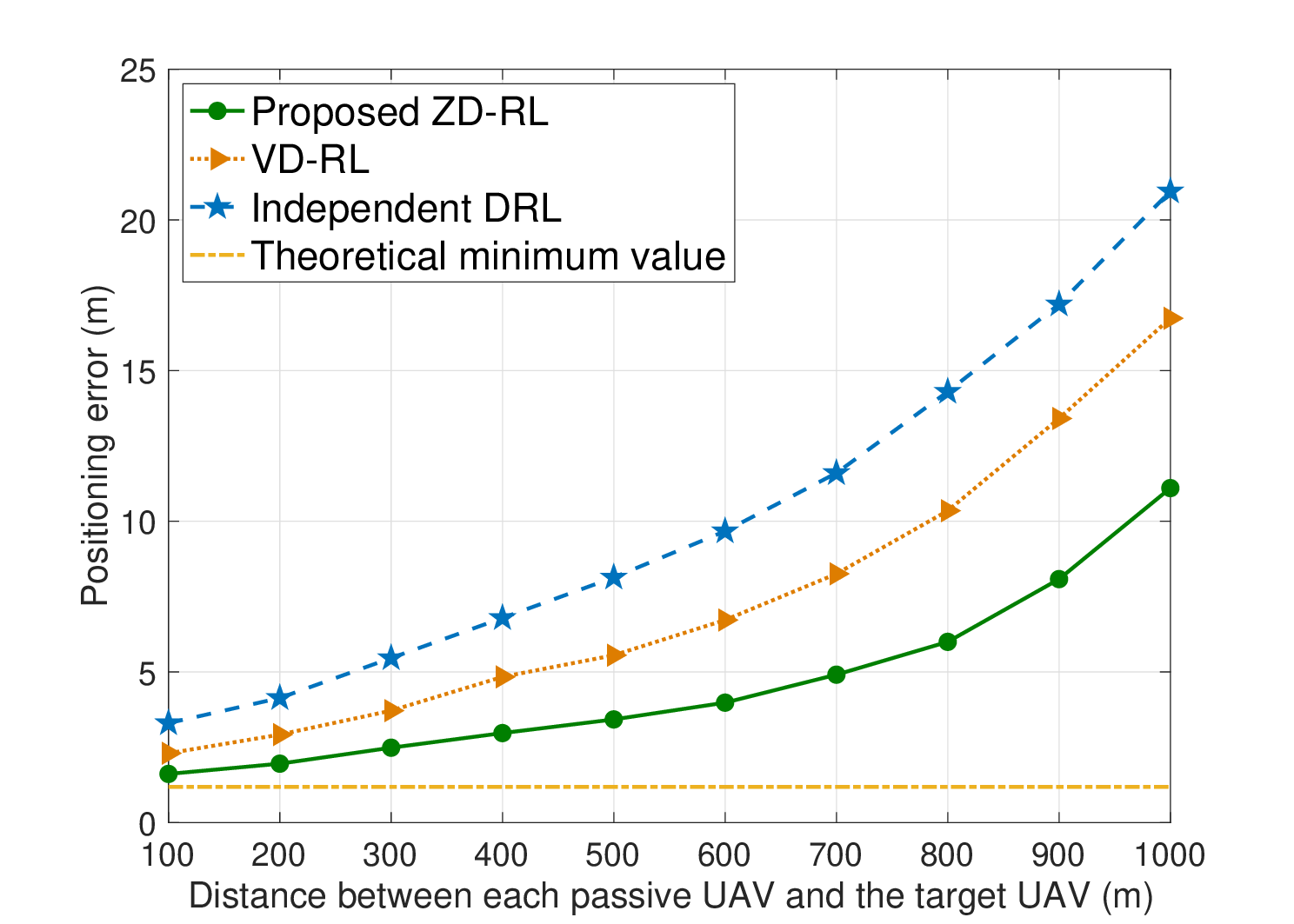}
\caption{Value of the positioning error as the distance between each controlled UAV and the target UAV varies.}
\label{d} 
\vspace{-0.6cm}
\end{figure} 

Fig. \ref{d} shows how the average positioning errors change as the distance between each controlled UAV and the target UAV varies. In this simulation, the target UAV moves in the curve of ``S'' and the distances between each controlled UAV and the target UAV satisfy $d_{1,t}=d_{2,t}=d_{3,t}=d_{4,t}$. The yellow line in Fig. \ref{d} represents the theoretically analytical result of the minimum positioning error obtained by Lemma 2. In Fig. \ref{d}, we can see that the minimum positioning error obtained by the proposed ZD-RL method is $1.61$ m while the theoretical positioning error is $1.18$ m when $d_{m,t}=$ 100 m. Hence, there is a gap between the theoretical and the simulation results. This is because the measurement information estimated by passive UAVs may have errors and the controlled UAVs may not be able to keep the minimum safety distance with the target UAV in real time. 
From Fig. \ref{d}, we can also see that the positioning errors of considered algorithms increase as the distance between each controlled UAV and the target UAV increases. This stems from the fact that the SNR of signals transmitted from the active UAV to each passive UAV via the target UAV decreases as the distance between each controlled UAV and the target UAV increase. Fig. \ref{d} also shows that the proposed ZD-RL method can reduce the positioning error by up to 33.6\% and 46.7\% compared to the VD-RL and independent DRL methods when $d_{m,t}=1000$ m. {This is because the proposed ZD-RL algorithm enables each controlled UAV to update its DNN parameters based on the approximated probability distribution of individual Z function and adjust its trajectory to minimize the positioning error of the target UAV cooperatively.} 

\begin{figure}[t!]
 \setlength\abovecaptionskip{0.cm}
\setlength\belowcaptionskip{-0.2cm}
\centering 
\includegraphics[width=8cm]{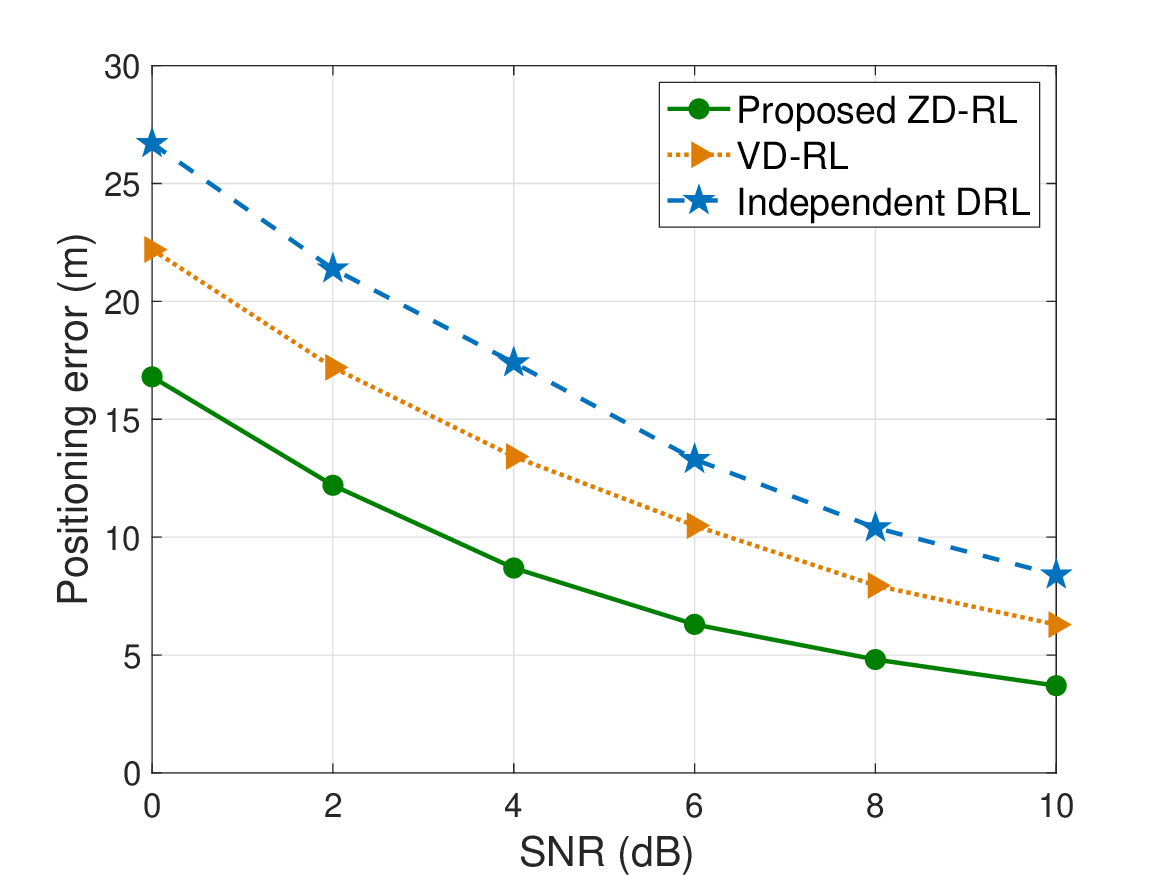}
\caption{Value of the positioning error as the SNR of signals transmitted from the target UAV to passive UAVs varies. ($d_{1,t}=d_{2,t}=d_{3,t}=d_{4,t}=900$ m)}
\label{SNR} 
\vspace{-0.6cm}
\end{figure} 
Fig. \ref{SNR} shows how the positioning errors change as the SNR of signals transmitted from the active UAV to each passive UAV varies. 
From Fig. \ref{SNR}, we can see that as SNR increases, the positioning errors obtained by considered algorithms decrease. This stems from the fact that the variance of measurement errors of each passive UAV increases as SNR decreases. Fig. \ref{SNR} also shows that the proposed algorithm can reduce positioning errors by up to 24.3\% and 37.1\% compared to VD-RL method and independent DRL method, respectively, when the SNR is 0 dB. This is because 
the proposed ZD-RL can approximate the expected value of the sum of future rewards using a non-linear weight function thus improve approximation accuracy. \textcolor{black}{From Fig. \ref{SNR}, we can see that as the SNR of each passive UAV increases, the positioning error of the target UAV decreases slowly. This is because the positioning accuracy of the target UAV is not only affected by SNRs of passive UAVs, but also the deployment of controlled UAVs. When SNR is small, the increase of SNR can significantly  decrease the positioning errors. However, as SNR continues to increases, the impact of SNR on positioning errors decreases and the deployment of controlled UAVs becomes the key factor that introduces of the positioning errors.}

\begin{figure}[t!]
 \setlength\abovecaptionskip{0.cm}
\setlength\belowcaptionskip{-0.2cm}
\centering
\includegraphics[width=8cm]{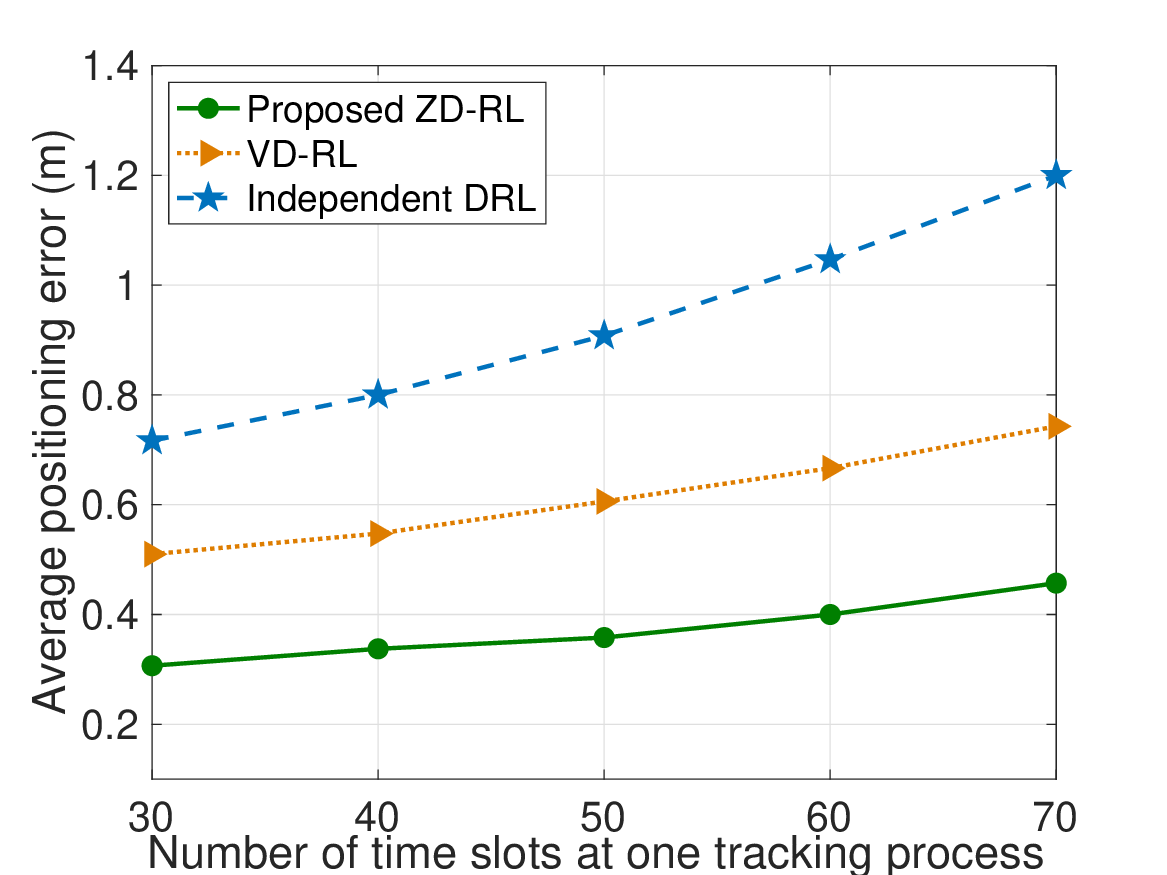}
\caption{\textcolor{black}{Average positioning error as the number of time slots at one tracking process varies.}}
\label{duration}
\vspace{-0.4cm}
\end{figure}

\textcolor{black}{Fig. \ref{duration} shows how the average positioning error $\Bar{e}_t=\frac{1}{V}\sum_{t=1}^V\sqrt{\left(\boldsymbol{s}_t-\hat{\boldsymbol{s}_t}\right)^2}$ of the target UAV changes as the number of time slots $V$ at one tracking process varies. From Fig. \ref{duration}, we see that when $V$ increases, the average positioning error of the ZD-RL increases slower compared to VD-RL and independent DRL methods. This is because the ZD-RL method can approximate the probability distribution of the sum of future rewards and capture richer information of the environment, thus estimating the expected value of the sum of rewards under selected actions more accurately compared to the VD-RL and independent DRL methods and optimally adjusting UAV trajectories to reduce the average positioning error.}

\begin{figure}[t!]
 \setlength\abovecaptionskip{0.cm}
\setlength\belowcaptionskip{-0.2cm}
\centering 
\includegraphics[width=8cm]{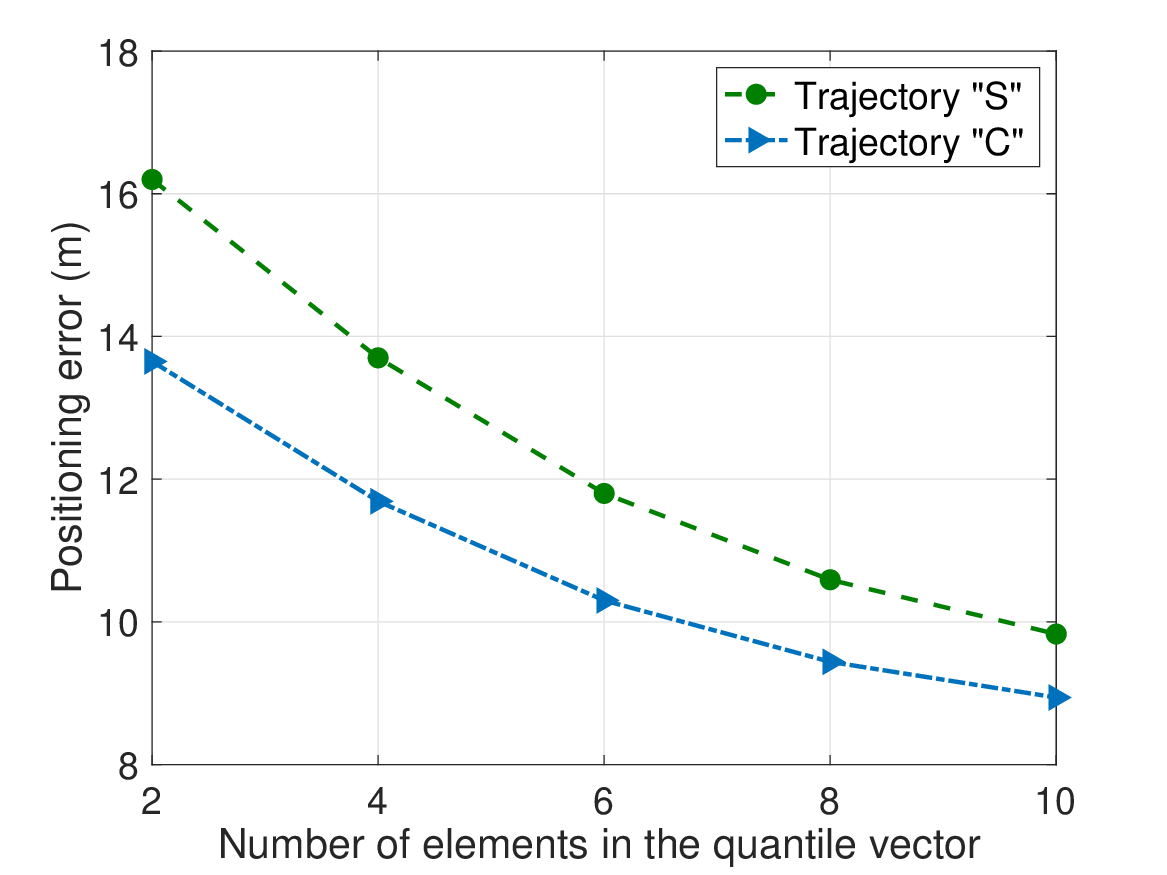}
\caption{Value of the positioning error as the number of elements $N$ in the quantile vector varies when the target UAV moves in the curve of ``S'' and ``C''.}
\label{N} 
\vspace{-0.6cm}
\end{figure} 

\begin{figure*}[h!]
 \setlength\abovecaptionskip{0.cm}
\setlength\belowcaptionskip{-0.2cm}
\centering
\subfigure[]{
\includegraphics[width=0.3\linewidth]{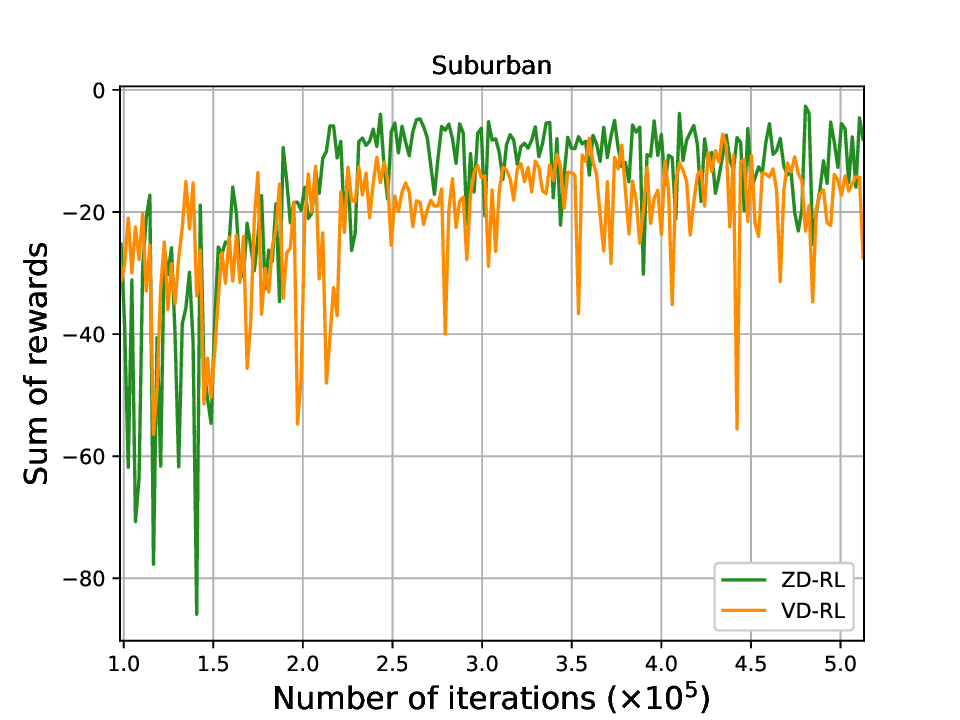}
}
\subfigure[]{
\includegraphics[width=0.3\linewidth]{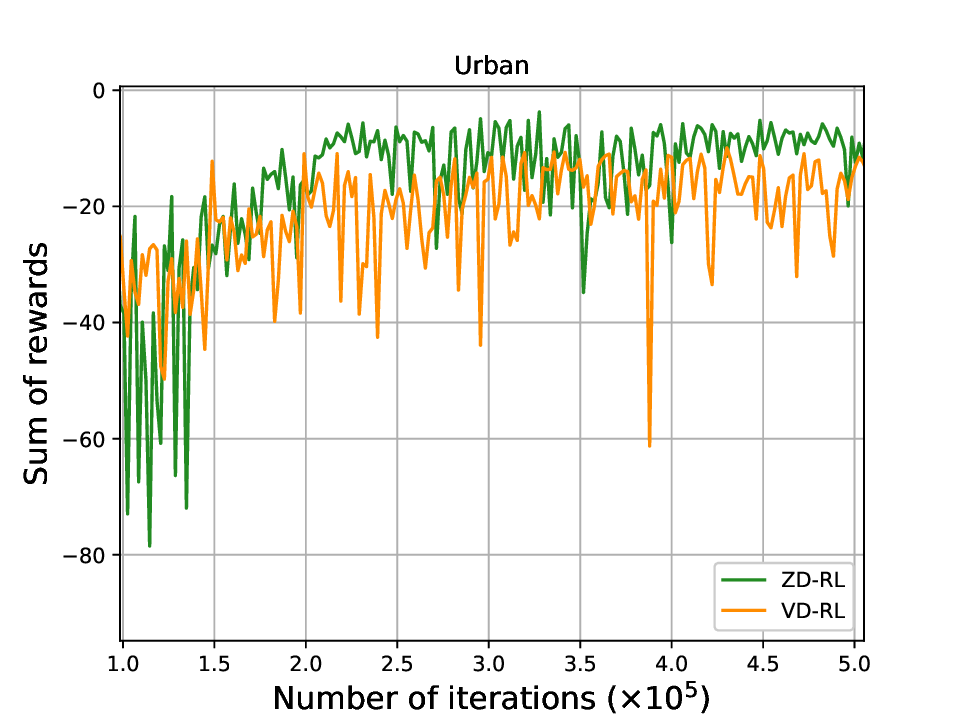}
}
\subfigure[]{
\includegraphics[width=0.3\linewidth]{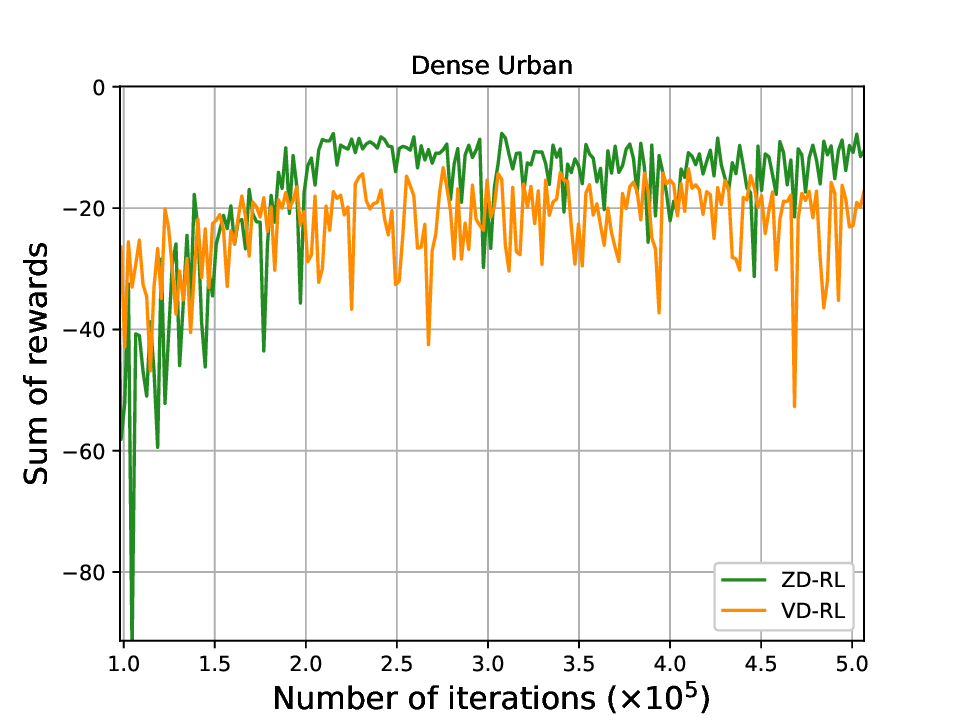}
}
\caption{\textcolor{black}{The sum of rewards as the number of iterations varies in different scenarios.}}
\label{scenario}
\vspace{-0.6cm}
\end{figure*}

Fig. \ref{N} shows how the positioning errors obtained by the proposed ZD-RL method change as the number of elements $N$ in the quantile vector varies.
From Fig. \ref{N}, we can see that as the value of $N$ increases, the positioning errors obtained by the proposed ZD-RL method decrease. This stems from the fact that when the number of elements in the quantile vector increases, each agent can obtain more values of the sum of future rewards with different quantiles thus approximating the probability distribution of individual Z functions more accurately.
Fig. \ref{N} also shows that the positioning error first drops rapidly when the number of quantiles is small and then decreases more slowly as the number of quantiles increases sufficiently. This is because as the number of quantiles is quite small, the localization performance is mainly limited by the fact that the proposed algorithm cannot accurately approximate the probability distribution of individual Z functions. When $N$ gradually increases, the main limitation shifts from the number of quantiles to the trajectory of the target UAV.

\textcolor{black}{Fig. \ref{scenario} shows how the sum of rewards obtained by the ZD-RL and VD-RL methods change as the number of iterations varies under different environments (Suburban, Urban, and Dense Urban \cite{channel}), in which the channel conditions are listed in Table \ref{channelconditions}. Figs. \ref{scenario}(a), \ref{scenario}(b), and \ref{scenario}(c) show the sum of rewards obtained by the ZD-RL and VD-RL methods under these scenarios. From Fig. \ref{scenario}, we see that the ZD-RL can obtain better localization performance than the VD-RL method in different environments. This is because the ZD-RL calculates the positioning error more accurately compared to the VD-RL method in different environments and optimally adjusts the trajectories of controlled UAVs.}

\begin{table}[t]
\caption{\label{channelconditions}\textcolor{black}{Channel Conditions} } 
\centering 
\setlength{\abovecaptionskip}{0.cm}
\setlength{\tabcolsep}{0.8mm}{
\begin{tabular}{|c|c|c|c|c|}  
\hline  
Scenarios & Suburban & Urban & Dense Urban \\ 
\hline  
$\left(\lambda_{\sigma_{\textrm{LoS}}},\lambda_{\sigma_{\textrm{NLoS}}}\right)$ & (0.1, 21)& (1.0, 20)& (1.6, 23)\\ 
\hline 
\end{tabular}}
\vspace{-0.5cm}
\end{table}

\begin{figure}[t!]
\setlength\abovecaptionskip{0.cm}
\setlength\belowcaptionskip{-0.2cm}
\centering
\includegraphics[width=8cm]{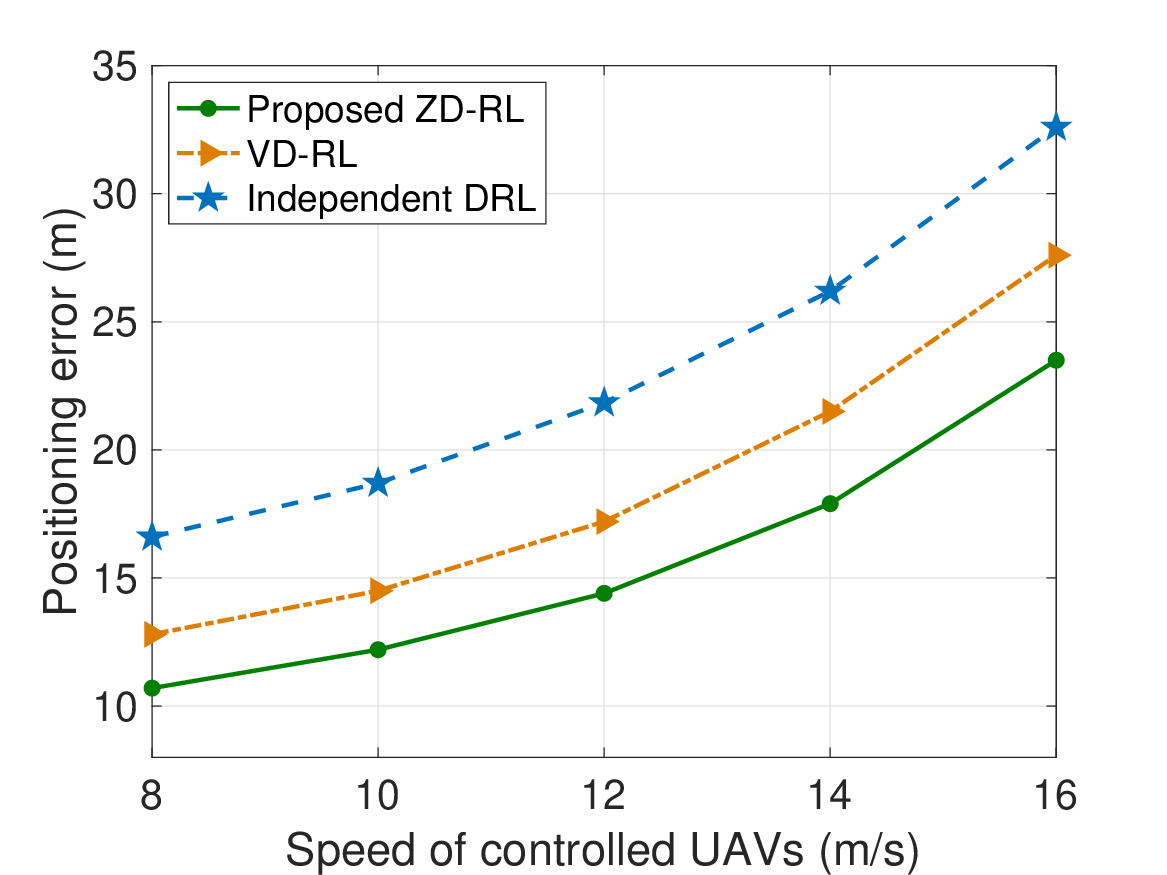}
\caption{\textcolor{black}{Positioning error as the speed of controlled UAVs varies under UAV flight energy consumption constraint.}}
\label{consumption}
\vspace{-0.6cm}
\end{figure}

\textcolor{black}{Since limited UAV flight energy affects the UAV trajectory optimization \cite{addadd3}, we analyze the localization performance of the ZD-RL method under limited UAV flight energy consumption constraint. We first model the flight energy consumption $E_{m,t}^\textrm{F}\left(\phi_{m,t}\right)$ of controlled UAV $m$ at time slot $t$ as \cite{flightenergy}
\begin{equation}
    \begin{split}    E_{m,t}^\textrm{F}\left(\phi_{m,t}\right)=&\frac{C_1\Delta_t}{\sqrt{\left(v_{m,t}^\textrm{L}\right)^2+\sqrt{\left(v_{m,t}^\textrm{L}\right)^4+4\left(v_{m,t}^\textrm{H}\right)^4}}}\\&+Mgv_{m,t}\sin{\phi_{m,t}}+C_2\left(v_{m,t}^\textrm{L}\right)^3,\end{split}
\end{equation}where $C_1$ and $C_2$ are coefficients \cite{flightenergy}, $v_{m,t}^\textrm{L}=v_{m,t}\cos{\phi_{m,t}}$ is the horizontal flight speed, $M$ is the weight of each controlled UAV, $g$ is the acceleration of gravity, and $v_{m,t}^\textrm{H}$ is the power needed for hovering. 
Then, under the flight energy consumption constraint $E_{m,t}^\textrm{F}\leqslant$ 500 J, Fig. \ref{consumption} shows how the positioning error of the target UAV changes as the speed of controlled UAVs varies under the maximal flight energy consumption constraint when the target UAV moves in the curve `C'. From Fig. \ref{consumption}, we see that the positioning errors obtained by the considered methods increase as the speed of controlled UAVs increases. This stems from the fact that the UAV flight energy consumption is proportional to the speed of controlled UAVs. Thus, the increase of the UAV's speed limits the UAV movement and increases the positioning error of the target UAV. Fig. \ref{consumption} also shows that the proposed ZD-RL can reduce the positioning error of the target UAV by up to 15.8\% and 34.7\% compared to VD-RL and independent DRL methods when the speed of controlled UAVs is 10 m/s. This is because the ZD-RL can estimate the sum of future rewards more accurately and thus can optimally adjust the trajectories of controlled UAVs to localize the target UAV under the energy consumption constraint.}

\begin{figure}[t!]
 \setlength\abovecaptionskip{0.cm}
\setlength\belowcaptionskip{-0.2cm}
\centering 
\includegraphics[width=8cm]{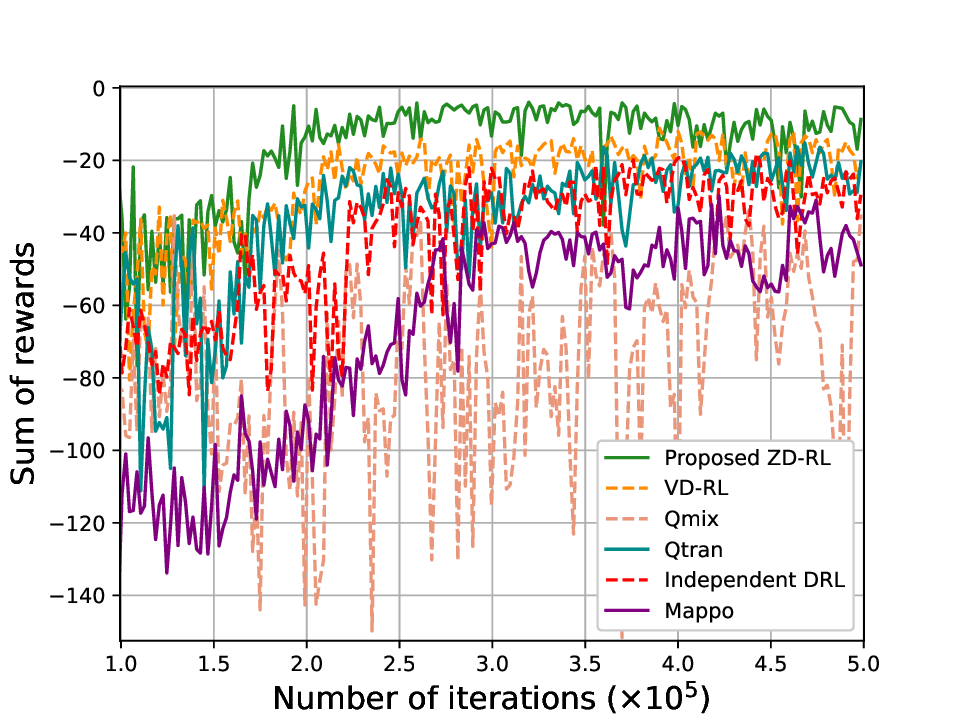}
\caption{\textcolor{black}{Value of the sum of rewards as the total number of iterations varies.}}
\label{reward} 
\vspace{-0.6cm}
\end{figure}

\textcolor{black}{ 
Fig. \ref{reward} shows how the positioning accuracy changes as the number of iterations varies. In this figure, we compare the proposed method with three other methods:   
1) Qmix method in which the BS uses a mixing network to combine individual Q function values of each controlled UAV into a global Q function value \cite{qmix}, 2) Qtran method that optimizes UAV trajectories by transforming actions of controlled UAVs into variables related to individual Q functions \cite{qtran}, and 3) Mappo method in which each controlled UAV optimizes its trajectory and controlled UAVs share agents' experiences \cite{Mappo,addadd4,addadd5}. 
From Fig. \ref{reward}, we see that the proposed ZD-RL method can improve the sum of rewards by up to 39.4\%, 54.6\%, 64.6\%, and 72.9\% compared to the VD-RL, Qtran, independent DRL, and Mappo methods, respectively. This stems from the fact that the ZD-RL method can approximate the probability distribution of the sum of discounted future rewards to calculate the expected value of the sum of future rewards more accurately compared to other baseline methods that estimate the expected value of the sum of future rewards directly.
Fig. \ref{reward} also shows that the proposed ZD-RL method can reduce the number of iterations required to converge by up to 9.0\%, 12.7\%, 19.35\%, and 30.8\% compared to the VD-RL, Qtran, independent DRL, and Mappo methods. The reason is that the proposed method cooperatively train the trajectories of controlled UAVs and the transmit power of the active UAV using the probability distribution of the sum of future rewards. Compared to other baselines that estimate the expected value of the sum of future rewards, the proposed ZD-RL method are more stable and accurate thus reducing the number of iterations required to convergence.
In particular, the number of iterations of the considered methods to converge is shown in Fig. \ref{reward} and the tested implementation time per iteration of each method is listed in Table \ref{complexity}. The total training times of the ZD-RL, VD-RL, Qtran, independent DRL, and Mappo methods to reach convergence are 1627.2 s, 1794.5 s, 1723.8 s, 1816.0 s, and 4436.4 s. Consequently, the ZD-RL can reduce the training complexity by up to 9.3\%, 5.6\%, 10.4\%, and 63.3\% compared to VD-RL, Qtran, independent DRL, and Mappo methods.}
\section{Conclusion}
In this paper, a novel localization framework that uses several controlled UAVs to localize a target UAV has been proposed. We have modeled this localization problem as an optimization problem that aims to optimize the positioning accuracy by jointly optimizing the transmit power of the active UAV and trajectories of all controlled UAVs. To solve this problem, we have proposed a ZD-RL method, which uses the probability distribution of the sum of future rewards to estimate the expected values of the sum of future rewards instead of directly estimating the expected values of the sum of future rewards as done in Deep Q. Hence, the proposed method enables each controlled UAV to find its optimal transmit power and trajectory to minimize the positioning errors efficiently. To further reduce the positioning error of the target UAV, we have derived the relationship between the positions of controlled UAVs and the positioning error of the target UAV. Based on the derived expression of the positioning error, we can obtain the minimum positioning error of the target UAV. Simulation results have shown that the proposed method yielded significant improvements in terms of the positioning accuracy compared to baselines.

\section*{Appendix}
\subsection{Proof of Lemma 1}
We first explain why the proposed ZD-RL method satisfies condition 1). From \eqref{loss}, the update rule of individual Z function of controlled UAV $m$ can be given by
\begin{equation}
\begin{split}
    Z_{k+1}\left(\boldsymbol{o}_{m,t},\boldsymbol{a}_{m,t}\right)&= Z_{k}\left(\boldsymbol{o}_{m,t},\boldsymbol{a}_{m,t}\right)+\alpha_m\left(R\left(\boldsymbol{o}_{m,t},\boldsymbol{a}_{m,t}\right)\right.\\&\left. +Z\left(\boldsymbol{o}_{m,t+1},\boldsymbol{a}_{m,t+1}\right)-Z\left(\boldsymbol{o}_{m,t},\boldsymbol{a}_{m,t}\right)\right).\end{split}
\end{equation} Taking the expectation of individual Z function with respect to transition probability distribution $\mathbb{P}\left(\boldsymbol{o}'_{m}|\boldsymbol{o}_{m,t},\boldsymbol{a}_{m,t}\right)$ and subtracting $M^*\left(\boldsymbol{o}_{m,t},\boldsymbol{a}_{m,t}\right)$ at both sides, we have
\begin{equation}\label{e1}
    \begin{split}
        \mathbb{E}&\left[Z_{k+1}\left(\boldsymbol{o}_{m,t},\boldsymbol{a}_{m,t}\right)\right]-M^*\left(\boldsymbol{o}_{m,t},\boldsymbol{a}_{m,t}\right)=\\&\left(1-\alpha_m\right)\left(\mathbb{E}\left[Z\left(\boldsymbol{o}_{m,t},\boldsymbol{a}_{m,t}\right)\right]-M^*\left(\boldsymbol{o}_{m,t},\boldsymbol{a}_{m,t}\right)\right)\\&+\alpha_m\left(R\left(\boldsymbol{o}_{m,t},\boldsymbol{a}_{m,t}\right)+\gamma \mathbb{E}\left[Z\left(\boldsymbol{o}_{m,t+1},\boldsymbol{a}_{m,t+1}\right)\right]\right.\\&\left. -M^*\left(\boldsymbol{o}_{m,t},\boldsymbol{a}_{m,t}\right)\right).
    \end{split}
\end{equation}
Since $e\left(\boldsymbol{o}_{m,t},\boldsymbol{a}_{m,t}\right)=M\left(\boldsymbol{o}_{m,t},\boldsymbol{a}_{m,t}\right)-M^*\left(\boldsymbol{o}_{m,t},\boldsymbol{a}_{m,t}\right)$ and $F\left(\boldsymbol{o}_{m,t},\boldsymbol{a}_{m,t}\right)=R\left(\boldsymbol{o}_{m,t},\boldsymbol{a}_{m,t}\right)+\gamma M\left(\boldsymbol{o}_{m,t+1},\boldsymbol{a}_{m,t+1}\right)-M^*\left(\boldsymbol{o}_{m,t},\boldsymbol{a}_{m,t}\right)$, we have
\begin{equation}    \begin{split}e_{k+1}&\left(\boldsymbol{o}_{m,t},\boldsymbol{a}_{m,t}\right)\\=&\left(1-\alpha_m\right)e_k\left(\boldsymbol{o}_{m,t},\boldsymbol{a}_{m,t}\right)+\alpha_m F\left(\boldsymbol{o}_{m,t},\boldsymbol{a}_{m,t}\right).\end{split}
\end{equation} Hence, the proposed ZD-RL method satisfies condition 1). Next, we explain why the proposed ZD-RL method satisfies condition 2). To prove condition 2), we first find the expected value of $F\left(\boldsymbol{o}_{m,t},\boldsymbol{a}_{m,t}\right)$, which is given by
\begin{equation}\label{e2}
    \begin{split}
    \mathbb{E}&\left[F\left(\boldsymbol{o}_{m,t},\boldsymbol{a}_{m,t}\right)\right]\\=&\mathbb{E}\left[R\left(\boldsymbol{o}_{m,t},\boldsymbol{a}_{m,t}\right)+\gamma M\left(\boldsymbol{o}_{m,t+1},\boldsymbol{a}_{m,t+1}\right)\right.\\&\left. -M^*\left(\boldsymbol{o}_{m,t},\boldsymbol{a}_{m,t}\right)\right]\\
    =&\mathbb{E}\left[R\left(\boldsymbol{o}_{m,t},\boldsymbol{a}_{m,t}\right)+\gamma \mathbb{E}\left[Z\left(\boldsymbol{o}_{m,t+1},\boldsymbol{a}_{m,t+1}\right)\right]\right]\\&-\mathbb{E}\left[Z^*\left(\boldsymbol{o}_{m,t},\boldsymbol{a}_{m,t}\right)\right]\\\overset{(a)}{=}&\mathbb{E}\left[\mathcal{T}\left(Z\left(\boldsymbol{o}_{m,t},\boldsymbol{a}_{m,t}\right)\right)\right]-\mathbb{E}\left[\mathcal{T}\left(Z^*\left(\boldsymbol{o}_{m,t},\boldsymbol{a}_{m,t}\right)\right)\right]\\\overset{(b)}{=}&\mathcal{T}\left(\mathbb{E}\left[Z\left(\boldsymbol{o}_{m,t},\boldsymbol{a}_{m,t}\right)\right]\right)-\mathcal{T}\left(\mathbb{E}\left[Z^*\left(\boldsymbol{o}_{m,t},\boldsymbol{a}_{m,t}\right)\right]\right),
    \end{split}
\end{equation} where equation (a) and equation (b) follow from the results in [\citenum{pth}, Lemma 4]. According to the results in [\citenum{pth}, Lemma 3], we have \begin{equation}\label{e3}
\begin{split}&\lvert\lvert\mathcal{T}\left(\mathbb{E}\left[Z\left(\boldsymbol{o}_{m,t},\boldsymbol{a}_{m,t}\right)\right]\right)-\mathcal{T}\left(\mathbb{E}\left[Z^*\left(\boldsymbol{o}_{m,t},\boldsymbol{a}_{m,t}\right)\right]\right)\rvert\rvert_{\infty}\\&\leqslant \gamma\lvert\lvert\mathbb{E}\left[Z\left(\boldsymbol{o}_{m,t},\boldsymbol{a}_{m,t}\right)\right]-\mathbb{E}\left[Z^*\left(\boldsymbol{o}_{m,t},\boldsymbol{a}_{m,t}\right)\right]\rvert\rvert_{\infty}.\end{split}
\end{equation} Based on \eqref{e3}, \eqref{e2} can be written as
\begin{equation}
    \begin{split}
    &\lvert\lvert\mathbb{E}\left[F\left(\boldsymbol{o}_{m,t},\boldsymbol{a}_{m,t}\right)\right]\rvert\rvert_{\infty}\\&=\lvert\lvert\mathcal{T}\left(\mathbb{E}\left[Z\left(\boldsymbol{o}_{m,t},\boldsymbol{a}_{m,t}\right)\right]\right)-\mathcal{T}\left(\mathbb{E}\left[Z^*\left(\boldsymbol{o}_{m,t},\boldsymbol{a}_{m,t}\right)\right]\right)\rvert\rvert_{\infty}\\&\leqslant \gamma\lvert\lvert\mathbb{E}\left[Z\left(\boldsymbol{o}_{m,t},\boldsymbol{a}_{m,t}\right)\right]-\mathbb{E}\left[Z^*\left(\boldsymbol{o}_{m,t},\boldsymbol{a}_{m,t}\right)\right]\rvert\rvert_{\infty}\\&=\gamma\lvert\lvert M\left(\boldsymbol{o}_{m,t},\boldsymbol{a}_{m,t}\right)-M^*\left(\boldsymbol{o}_{m,t},\boldsymbol{a}_{m,t}\right)\rvert\rvert_{\infty}\\&=\gamma\lvert\lvert e\left(\boldsymbol{o}_{m,t},\boldsymbol{a}_{m,t}\right)\rvert\rvert_{\infty}.
    \end{split}
\end{equation} Hence, condition 2) is satisfied. For condition 3), using \eqref{e2}, $\mathrm{Var}\left(\mathbb{E}\left[F\left(\boldsymbol{o}_{m,t},\boldsymbol{a}_{m,t}\right)\right]\right)$ can be rewritten as
\begin{equation}\label{e5}
    \begin{split}
\mathrm{Var}&\left(\mathbb{E}\left[F\left(\boldsymbol{o}_{m,t},\boldsymbol{a}_{m,t}\right)\right]\right)\\=&\mathbb{E}\left[\left(F\left(\boldsymbol{o}_{m,t},\boldsymbol{a}_{m,t}\right)-\mathbb{E}\left[F\left(\boldsymbol{o}_{m,t},\boldsymbol{a}_{m,t}\right)\right]\right)^2\right]\\=&\mathbb{E}\left[F\left(\boldsymbol{o}_{m,t},\boldsymbol{a}_{m,t}\right)\right.\\&\left.-\left(\mathcal{T}\left(M\left(\boldsymbol{o}_{m,t},\boldsymbol{a}_{m,t}\right)\right)-\mathcal{T}\left(M^*\left(\boldsymbol{o}_{m,t},\boldsymbol{a}_{m,t}\right)\right)\right)^2\right]\\=&\mathbb{E}\left[R\left(\boldsymbol{o}_{m,t},\boldsymbol{a}_{m,t}\right)+\gamma M\left(\boldsymbol{o}_{m,t+1},\boldsymbol{a}_{m,t+1}\right)\right.\\&\left.-\left(R\left(\boldsymbol{o}_{m,t},\boldsymbol{a}_{m,t}\right)+\gamma \mathbb{E}\left[M\left(\boldsymbol{o}_{m,t},\boldsymbol{a}_{m,t}\right)\right]\right)^2\right]\\=&\gamma^2\mathbb{E}\left[\left(M\left(\boldsymbol{o}_{m,t+1},\boldsymbol{a}_{m,t+1}\right)-\mathbb{E}\left[M\left(\boldsymbol{o}_{m,t},\boldsymbol{a}_{m,t}\right)\right]\right)^2\right]\\=&\gamma^2\mathrm{Var}\left(\mathbb{E}\left[M\left(\boldsymbol{o}_{m,t+1},\boldsymbol{a}_{m,t+1}\right)\right]\right)\\\leqslant&\gamma^2\mathbb{E}\left[M\left(\boldsymbol{o}_{m,t+1},\boldsymbol{a}_{m,t+1}\right)^2\right]\\\leqslant&\gamma^2\max_{\boldsymbol{o}_{m,t}},\max_{\boldsymbol{a}_{m,t}}\left(M\left(\boldsymbol{o}_{m,t},\boldsymbol{a}_{m,t}\right)\right)^2
    \\\leqslant&\gamma^2\lvert\lvert e\left(\boldsymbol{o}_{m,t},\boldsymbol{a}_{m,t}\right)+M^*\left(\boldsymbol{o}_{m,t},\boldsymbol{a}_{m,t}\right)\rvert\rvert_{\infty}^2\\=&\gamma^2\lvert\lvert e\left(\boldsymbol{o}_{m,t},\boldsymbol{a}_{m,t}\right)\rvert\rvert_{\infty}^2+\gamma^2\lvert\lvert M^*\left(\boldsymbol{o}_{m,t},\boldsymbol{a}_{m,t}\right)\rvert\rvert_{\infty}^2\\&+2\gamma^2\lvert\lvert e\left(\boldsymbol{o}_{m,t},\boldsymbol{a}_{m,t}\right)\rvert\rvert_{\infty}\lvert\lvert M^*\left(\boldsymbol{o}_{m,t},\boldsymbol{a}_{m,t}\right)\rvert\rvert_{\infty}.
    \end{split}
\end{equation}Since the value of $\mathrm{Var}\left(\mathbb{E}\left[F\left(\boldsymbol{o}_{m,t},\boldsymbol{a}_{m,t}\right)\right]\right)$ depends on $\lvert\lvert e\left(\boldsymbol{o}_{m,t},\boldsymbol{a}_{m,t}\right)\rvert\rvert_{\infty}$, next, we calculate the maximum value of $\mathrm{Var}\left(\mathbb{E}\left[F\left(\boldsymbol{o}_{m,t},\boldsymbol{a}_{m,t}\right)\right]\right)$ according to the value of $\lvert\lvert e\left(\boldsymbol{o}_{m,t},\boldsymbol{a}_{m,t}\right)\rvert\rvert\leqslant 1$. In particular, when $\lvert\lvert e\left(\boldsymbol{o}_{m,t},\boldsymbol{a}_{m,t}\right)\rvert\rvert_{\infty}\leqslant 1$, \eqref{e5} can be written as
\begin{equation}\label{e6}
    \begin{split}
        \gamma^2&\lvert\lvert e\left(\boldsymbol{o}_{m,t},\boldsymbol{a}_{m,t}\right)\rvert\rvert_{\infty}^2+\gamma^2\lvert\lvert M^*\left(\boldsymbol{o}_{m,t},\boldsymbol{a}_{m,t}\right)\rvert\rvert_{\infty}^2\\&+2\gamma^2\lvert\lvert e\left(\boldsymbol{o}_{m,t},\boldsymbol{a}_{m,t}\right)\rvert\rvert_{\infty}\lvert\lvert M^*\left(\boldsymbol{o}_{m,t},\boldsymbol{a}_{m,t}\right)\rvert\rvert_{\infty}\\\leqslant &\gamma^2\lvert\lvert e\left(\boldsymbol{o}_{m,t},\boldsymbol{a}_{m,t}\right)\rvert\rvert_{\infty}^2+2\gamma^2\lvert\lvert M^*\left(\boldsymbol{o}_{m,t},\boldsymbol{a}_{m,t}\right)\rvert\rvert_{\infty}\\&+\gamma^2\lvert\lvert M^*\left(\boldsymbol{o}_{m,t},\boldsymbol{a}_{m,t}\right)\rvert\rvert_{\infty}^2\\\leqslant&\gamma^2\left(\lvert\lvert M^*\left(\boldsymbol{o}_{m,t},\boldsymbol{a}_{m,t}\right)\rvert\rvert_{\infty}^2+2\lvert\lvert M^*\left(\boldsymbol{o}_{m,t},\boldsymbol{a}_{m,t}\right)\rvert\rvert_{\infty}\right)\\&\times\left(1+\lvert\lvert e\left(\boldsymbol{o}_{m,t},\boldsymbol{a}_{m,t}\right)\rvert\rvert^2_{\infty}\right).
    \end{split}
\end{equation} If $\lvert\lvert e\left(\boldsymbol{o}_{m,t},\boldsymbol{a}_{m,t}\right)\rvert\rvert_{\infty}\geqslant 1$, we have $\lvert\lvert e\left(\boldsymbol{o}_{m,t},\boldsymbol{a}_{m,t}\right)\rvert\rvert_{\infty}\leqslant\lvert\lvert e\left(\boldsymbol{o}_{m,t},\boldsymbol{a}_{m,t}\right)\rvert\rvert_{\infty}^2$ and \eqref{e5} can be rewritten as
\begin{equation}\label{e7}
    \begin{split}
        &\gamma^2\lvert\lvert e\left(\boldsymbol{o}_{m,t},\boldsymbol{a}_{m,t}\right)\rvert\rvert_{\infty}^2+\gamma^2\lvert\lvert M^*\left(\boldsymbol{o}_{m,t},\boldsymbol{a}_{m,t}\right)\rvert\rvert_{\infty}^2\\&+2\gamma^2\lvert\lvert e\left(\boldsymbol{o}_{m,t},\boldsymbol{a}_{m,t}\right)\rvert\rvert_{\infty}\lvert\lvert M^*\left(\boldsymbol{o}_{m,t},\boldsymbol{a}_{m,t}\right)\rvert\rvert_{\infty}\\\leqslant& \gamma^2\left(1+2\lvert\lvert M^*\left(\boldsymbol{o}_{m,t},\boldsymbol{a}_{m,t}\right)\rvert\rvert_{\infty}\right)\lvert\lvert e\left(\boldsymbol{o}_{m,t},\boldsymbol{a}_{m,t}\right)\rvert\rvert_{\infty}^2\\&+\gamma^2\lvert\lvert M^*\left(\boldsymbol{o}_{m,t},\boldsymbol{a}_{m,t}\right)\rvert\rvert_{\infty}^2\\\leqslant& \gamma^2\left(1+2\lvert\lvert M^*\left(\boldsymbol{o}_{m,t},\boldsymbol{a}_{m,t}\right)\rvert\rvert_{\infty}\right)\left(1+\lvert\lvert e\left(\boldsymbol{o}_{m,t},\boldsymbol{a}_{m,t}\right)\rvert\rvert_{\infty}^2\right).
    \end{split}
\end{equation} Based on \eqref{e6} and \eqref{e7}, we have 
\begin{equation}
\mathrm{Var}\left(\mathbb{E}\left[F\left(\boldsymbol{o}_{m,t},\boldsymbol{a}_{m,t}\right)\right]\right)\leqslant C_{\textrm{F}}\left(1+\lvert\lvert e\left(\boldsymbol{o}_{m,t},\boldsymbol{a}_{m,t}\right)\rvert\rvert^2_{\infty}\right),
\end{equation} where $C_{\textrm{F}}$ is the maximal value of $2\gamma^2\lvert\lvert M^*\left(\boldsymbol{o}_{m,t}\boldsymbol{a}_{m,t}\right)\rvert\rvert_{\infty}+\gamma^2\lvert\lvert M^*\left(\boldsymbol{o}_{m,t},\boldsymbol{a}_{m,t}\right)\rvert\rvert_{\infty}^2$ and $\gamma^2\left(1+2\lvert\lvert M^*\left(\boldsymbol{o}_{m,t},\boldsymbol{a}_{m,t}\right)\rvert\rvert_{\infty}\right)$. Hence, condition 3) is satisfied. 
This completes the proof.
\subsection{Proof of Theorem 2}
Since $e_t=\sqrt{\textrm{tr}\left(\mathbb{E}\left[\textrm{d}\boldsymbol{s}_{t}\textrm{d}\boldsymbol{s}_{t}^{T}\right]\right)}$, we first calculate the value of $\mathbb{E}\left[\textrm{d}\boldsymbol{s}_{t}\textrm{d}\boldsymbol{s}_{t}^{T}\right]$. From \eqref{m5}, we have 
\begin{equation}\label{5203}    \textrm{d}\boldsymbol{s}_{t}=\left(\boldsymbol{M}^T\boldsymbol{M}\right)^{-1}\boldsymbol{M}^Td\boldsymbol{r}_t,
\end{equation} and the positioning error $e_t$ of the target UAV at time slot $t$ can be rewritten as
 \begin{equation}\label{m11}
     \begin{split}\mathbb{E}&\left[\textrm{d}\boldsymbol{s}_{t}\textrm{d}\boldsymbol{s}_{t}^{T}\right]\\&=\mathbb{E}\left[\left(\boldsymbol{M}^T\boldsymbol{M}\right)^{-1}\boldsymbol{M}^T\textrm{d}\boldsymbol{r}_t\left(\left(\boldsymbol{M}^T\boldsymbol{M}\right)^{-1}\boldsymbol{M}^T\textrm{d}\boldsymbol{r}_t\right)^T\right]\\&=\mathbb{E}\left[\left(\boldsymbol{M}^T\boldsymbol{M}\right)^{-1}\boldsymbol{M}^T\textrm{d}\boldsymbol{r}_t\textrm{d}\boldsymbol{r}_t^T\left(\left(\boldsymbol{M}^T\boldsymbol{M}\right)^{-1}\boldsymbol{M}^T\right)^T\right]\\&=\left(\boldsymbol{M}^T\boldsymbol{M}\right)^{-1}\boldsymbol{M}^T\mathbb{E}\left[\textrm{d}\boldsymbol{r}_t\textrm{d}\boldsymbol{r}_t^T\right]\left(\left(\boldsymbol{M}^T\boldsymbol{M}\right)^{-1}\boldsymbol{M}^T\right)^T,
     \end{split}
 \end{equation}where $\boldsymbol{M}^T$ is a transpose matrix of $\boldsymbol{M}$, $\left(\boldsymbol{M}^T\boldsymbol{M}\right)^{-1}$ is an inverse matrix of $\boldsymbol{M}^T\boldsymbol{M}$, $\mathbb{E}\left[\textrm{d}\boldsymbol{r}_t\textrm{d}\boldsymbol{r}_t^T\right]=\textrm{diag}\left(\sigma_{1,t}^2,\sigma_{2,t}^2,\sigma_{3,t}^2,\sigma_{4,t}^2\right)$
with $\sigma_{m,t}^2=k\left(d_{m,t}+d_{0,t}\right)^2$ being the variance of the independent Gaussian measurement error of passive UAV $m$ at time slot $t$ and $k$ being a coefficient \cite{variance}. 
Since $d_{1,t}=d_{2,t}=d_{3,t}=d_{4,t}$, $\mathbb{E}\left[\textrm{d}\boldsymbol{r}_t\textrm{d}\boldsymbol{r}_t^T\right]$ can be rewritten as
\begin{equation}\label{m12}
    \mathbb{E}\left[\textrm{d}\boldsymbol{r}_t\textrm{d}\boldsymbol{r}_t^T\right]=k\left(d_{m,t}+d_{0,t}\right)^2\boldsymbol{I},
\end{equation} where $\boldsymbol{I}=\textrm{diag}\left(1,1,1,1\right)$. Substituting \eqref{m12} into \eqref{m11}, we have 
\begin{equation}\label{5201}
    \begin{split}
        &\mathbb{E}\left[\textrm{d}\boldsymbol{s}_{t}\textrm{d}\boldsymbol{s}_{t}^{T}\right]\\&=k\left(d_{m,t}+d_{0,t}\right)^2\left(\boldsymbol{M}^T\boldsymbol{M}\right)^{-1}\boldsymbol{M}^T\left(\left(\boldsymbol{M}^T\boldsymbol{M}\right)^{-1}\boldsymbol{M}^T\right)^T\\&=k\left(d_{m,t}+d_{0,t}\right)^2\left(\boldsymbol{M}^T\boldsymbol{M}\right)^{-1}\boldsymbol{M}^T\boldsymbol{M}\left(\boldsymbol{M}^T\boldsymbol{M}\right)^{-1}\\&=k\left(\boldsymbol{M}^T\boldsymbol{M}\right)^{-1}.
    \end{split}
\end{equation}

Based on \eqref{5201}, the positioning error $e_t$ of the target UAV can be given by
\begin{equation}\label{5211}
    \begin{split}
        e_t&=\sqrt{\textrm{tr}\left(k\left(d_{m,t}+d_{0,t}\right)^2\left(\boldsymbol{M}^T\boldsymbol{M}\right)^{-1}\right)}\\&=\sqrt{k\left(d_{m,t}+d_{0,t}\right)^2\textrm{tr}\left(\left(\boldsymbol{M}^T\boldsymbol{M}\right)^{-1}\right)}\\&\overset{(a)}{\geqslant}\sqrt{4kL_{\textrm{min}}^2\textrm{tr}\left(\left(\boldsymbol{M}^T\boldsymbol{M}\right)^{-1}\right)},
    \end{split}
\end{equation}
where equation (a) stems from the fact that the distance $d_{m,t}$ between each controlled UAV and the target UAV satisfy $d_{m,t}\geqslant L_{\textrm{min}}, m=0,\cdots,4$, according to constraint \eqref{TC}. Therefore, equation (a) is hold when $d_{m,t}=d_{0,t}=L_{\textrm{min}}$. This completes the proof.
\subsection{Proof of Lemma 2}
Given the positions $\boldsymbol{s}_t$ and $\boldsymbol{u}_{0,t}$, the distance $d_{0,t}$ between the target UAV and the active UAV is a constant and \eqref{m2} can be rewritten as  
\begin{equation}\label{5202}
 \begin{split}
     \small{
     \textrm{d}r_{m,t}=\frac{x_t-x_{m,t}}{d_{m,t}}\textrm{d}x_t+\frac{y_t-y_{m,t}}{d_{m,t}}\textrm{d}y_t+\frac{z_t-z_{m,t}}{d_{m,t}}\textrm{d}z_t.}
 \end{split}
 \end{equation}
Then, the value of $\boldsymbol{M}$ in Theorem 2 can be rewritten as
    \begin{equation}\label{5205}
 \boldsymbol{M}=\frac{1}{d_{m,t}}\begin{bmatrix}
         x_t-x_{1,t}&y_t-y_{1,t}&z_t-z_{1,t}\\
         x_t-x_{2,t}&y_t-y_{2,t}&z_t-z_{2,t}\\
         x_t-x_{3,t}&y_t-y_{3,t}&z_t-z_{3,t}\\
         x_t-x_{4,t}&y_t-y_{4,t}&z_t-z_{4,t}\\
     \end{bmatrix}.
 \end{equation}
From \eqref{5211}, the positioning error $e_t$ can be written as $e_t=\sqrt{k\left(d_{m,t}+d_{0,t}^2\right)\textrm{tr}\left(\left(\boldsymbol{M}^T\boldsymbol{M}\right)^{-1}\right)}$. 
Since $\textrm{tr}\left(\left(\boldsymbol{M}^T\boldsymbol{M}\right)^{-1}\right)=\sum_{i=1}^3 \frac{1}{\varrho_i}$ with $\varrho_i$ being the eigenvalue of $\boldsymbol{M}^T\boldsymbol{M}$ \cite{5290368}, $e_t$ can be rewritten as
\begin{equation}\label{5208}
    \begin{split}
        e_t&=\sqrt{k\left(d_{m,t}+d_{0,t}\right)^2\sum_{i=1}^3 \frac{1}{\varrho_i}}\\&\overset{(a)}{\geqslant} \sqrt{k\left(d_{m,t}+d_{0,t}\right)^23\left(\prod_{i=1}^3 \frac{1}{\varrho_i}\right)^{\frac{1}{3}}}\\&\overset{(b)}{=} \sqrt{k\left(d_{m,t}+d_{0,t}\right)^23\left(\frac{3}{\textrm{tr}\left(\boldsymbol{M}^T\boldsymbol{M}\right)}\right)},
    \end{split}
\end{equation}where equation (a) is achieved by the triangle-inequality and equation (a) is hold when $\varrho_1=\varrho_2=\varrho_3$, equation (b) stems from the fact that $\varrho_1+\varrho_2+\varrho_3=\textrm{tr}\left(\boldsymbol{M}^T\boldsymbol{M}\right)$ and $\varrho_i=\frac{1}{3}\textrm{tr}\left(\boldsymbol{M}^T\boldsymbol{M}\right)$ when $\varrho_1=\varrho_2=\varrho_3$. Based on \eqref{5205}, $\textrm{tr}\left(\boldsymbol{M}^T\boldsymbol{M}\right)$ is given by
\begin{equation}\label{5207}
    \begin{split}
        \textrm{tr}&\left(\boldsymbol{M}^T\boldsymbol{M}\right)\\=&\frac{1}{d_{m,t}^2}\left(\sum_{m=1}^4\left(x_t-x_{m,t}\right)^2+\sum_{m=1}^4\left(y_t-y_{m,t}\right)^2\right.\\&\left. +\sum_{m=1}^4\left(z_t-z_{m,t}\right)^2\right)\\=&\frac{1}{d_{m,t}^2}\sum_{m=1}^4\left(\left(x_t-x_{m,t}\right)^2+\left(y_t-y_{m,t}\right)^2+\left(z_t-z_{m,t}\right)^2\right)\\=&\frac{1}{d_{m,t}^2}\left(\sum_{m=1}^4d_{m,t}^2\right)\overset{(a)}{=}4,
    \end{split}
\end{equation}where equation (a) stems from the fact that $d_{1,t}=d_{2,t}=d_{3,t}=d_{4,t}$. Substituting \eqref{5207} into \eqref{5208}, we have
\begin{equation}
    \begin{split}
        e_t&=\sqrt{k\left(d_{m,t}+d_{0,t}\right)^23\left(\frac{3}{4}\right)}\\&=\sqrt{\frac{9}{4}k\left(d_{m,t}+d_{0,t}\right)^2}\\&\overset{(a)}{\geqslant}\frac{3}{2}\left(L_{\textrm{min}}+d_{0,t}\right)\sqrt{k},
    \end{split}
\end{equation}where equation (a) stems from the fact that $d_{m,t}\geqslant L_{\textrm{min}}$ as shown in \eqref{TC}. This completes the proof.

\footnotesize
\bibliographystyle{IEEEtran}
\bibliography{IEEEabrv,Temp_Journal}

\end{CJK}
\end{document}